# Strain-Induced Relaxor Multiferroicity at Room Temperature in Hexaferrite $BaFe_{12}O_{19}$ Thin Films

Yilin Evan Li[1*], Harikrishnan KP[2*], Sankalpa Hazra[3*], Zhiren He[2*], Jayanti Higgins[2], ChanJu You[4], Mario Brützam[5], Jiaqiang Yan[6,7], Anna Park[1], Maya Ramesh[1], Wenwen Zhao[2], Jayda Shine[1], David G. Mandrus[6,7], Ramamoorthy Ramesh[8,9], Ankit S. Disa[2], Christo Guguschev[5], Guru Khalsa[10], Craig J. Fennie[2,†], Venkatraman Gopalan[3,11], Yu-Tsun Shao[12], David A. Muller[2,13], Darrell G. Schlom[1,5,13]

[1] Department of Materials Science and Engineering, Cornell University, Ithaca, NY 14853, USA.

[2] School of Applied and Engineering Physics, Cornell University, Ithaca, NY 14853, USA.

[3] Department of Materials Science and Engineering, The Pennsylvania State University, University Park, PA 16802, USA.

[4] Department of Physics, Cornell University, Ithaca, NY 14853, USA.

[5] Leibniz-Institut für Kristallzüchtung, Berlin 12489, Germany.

[6] Materials Science and Technology Division, Oak Ridge National Laboratory, Oak Ridge, TN 37830, USA.

[7] Department of Materials Science and Engineering, University of Tennessee, Knoxville, TN 37916, USA.

[8] Department of Materials Science and Engineering, University of California, Berkeley, CA 94720, USA.

[9] Department of Physics, University of California, Berkeley, CA 94720, USA.

[10] Department of Physics, University of North Texas, Denton, TX 76203, USA.

[11] Department of Physics, The Pennsylvania State University, University Park, PA 16802, USA.

[12] Mork Family Department of Chemical Engineering and Materials Science, University of Southern California, Los Angeles, CA 90089, USA.

[13] Kavli Institute at Cornell for Nanoscale Science, Ithaca, NY 14853, USA.

* These authors contributed equally to this work

† Deceased

## Abstract

Multiferroic materials that combine magnetic and electric order at room temperature are rare. Here, we demonstrate strain-induced room-temperature polar order in the ferrimagnetic hexaferrite $BaFe_{12}O_{19}$. First-principles calculations reveal a strain-tunable energy landscape with multiple competing dipolar configurations and predict that compressive strain favors polar distortions. Using an isostructural $Sr_{1.03}Ga_{10.81}Mg_{0.58}Zr_{0.58}O_{19}$ substrate, we grow coherently strained $BaFe_{12}O_{19}$ films with 1.1% in-plane biaxial compression. Second-harmonic generation measurements demonstrate inversion-symmetry breaking and establish a strain-stabilized polar phase that persists to at least 1000 K. Multislice electron ptychography directly reveals enhanced off-centering of $Fe^{3+}$ ions within the trigonal-bipyramidal sites of the strained films and spatially varying local polarization, demonstrating the formation of polar nanoregions. Path-integral Monte Carlo simulations further show that compressive strain suppresses quantum fluctuations and stabilizes these local polar distortions. Together, these results establish strain-engineered $BaFe_{12}O_{19}$ as a room-temperature relaxor multiferroic, in which robust ferrimagnetism coexists with nanoscale polar order. Our work demonstrates a route for transforming an incipient ferroelectric ferrimagnetic into a polar magnetic material through epitaxial strain.

## Significance Statement

Single-phase multiferroic materials, which simultaneously exhibit magnetic and ferroelectric ordering at room temperature, are exceptionally rare in nature. Here, epitaxial strain engineering is used to induce room-temperature relaxor multiferroicity in $BaFe_{12}O_{19}$, a strong room-temperature ferrimagnet whose bulk ferroelectricity is normally suppressed by quantum fluctuations. By growing coherently strained thin films on an isostructural substrate, compressive strain breaks structural inversion symmetry and stabilizes polar nanoregions that persist above 1000 K. Verified through optical second harmonic generation, atomic-scale 3D electron ptychography, and simulations, this work demonstrates a generalizable strategy to transmute room-temperature magnets into multiferroics.

# 1 Introduction

Multiferroic materials, which simultaneously exhibit two or more ferroic orders, such as ferroelectricity and magnetism, are considered ideal candidates for energy-efficient memory and logic applications[1,2]. Despite decades of research, multiferroics with both ferroelectric and magnetic Curie temperatures above room temperature are exceedingly rare; $BiFeO_3$ remains the most extensively studied multiferroic[2,3], with only a few additional materials identified as room-temperature multiferroics[4-9].

Epitaxial strain has been widely utilized to enhance ferroic orders or even induce them, as demonstrated in various materials. For instance, either tensile or compressive strain in $SrTiO_3$, an incipient ferroelectric, stabilizes ferroelectricity that is absent in its bulk form[10,11]. The same holds for $KTaO_3$[12]. Similarly, biaxial compressive strain in $BaTiO_3$ and $KNbO_3$ significantly increases their ferroelectric Curie temperatures and remanent polarization[13,14]. In the case of $EuTiO_3$, biaxial tensile strain induces simultaneous ferroelectric and magnetic ordering, making it a multiferroic despite its bulk form being neither ferroelectric nor ferromagnetic[15]. Yet, its magnetic Curie temperature is extremely low (~4.2 K), precluding it from being a room-temperature multiferroic.

A promising approach to realizing room-temperature multiferroics is applying epitaxial strain to a robust room-temperature magnetic material that is also an incipient ferroelectric. Lattice distortion can break inversion symmetry and stabilize a polar phase, offering a viable pathway for designing multiferroics that could operate under technologically relevant conditions.

Epitaxially strained $BaFe_{12}O_{19}$ is a strong candidate for a single-phase room-temperature multiferroic. In its bulk form, $BaFe_{12}O_{19}$ is a robust ferrimagnet with a net magnetic moment of 20 μB per formula unit and a high Curie temperature of 723 K[16], making it widely used as permanent magnets. Long recognized as an incipient ferroelectric[17,18], $BaFe_{12}O_{19}$ is predicted to transform into a ferroelectric at room temperature when subjected to more than 5% in-plane biaxial compressive strain[19]. A major challenge in achieving epitaxial strain in $BaFe_{12}O_{19}$, however, is the lack of an isostructural substrate with a slightly smaller in-plane lattice constant to provide a coherent compressive strain. Previous attempts to impose epitaxial strain on 200 nm thick $BaFe_{12}O_{19}$ films using non-isostructural substrates with 6% lattice mismatch likely suffered from relaxation of the attempted strain during film growth, obscuring the effect of strain[20].

In this work, we perform density functional theory (DFT) calculations to examine the effect of strain, utilize a novel, non-commercial substrate to apply a 1.1% in-plane biaxial compressive strain to $BaFe_{12}O_{19}$ films up to 27.5 nm thick, and validate symmetry breaking and the presence of polar nanoregions using second-harmonic generation and multislice electron ptychography.

# 2 Theory Prediction of Strain-Induced Ferroelectricity in $BaFe_{12}O_{19}$

Strain engineering has been proposed as a promising route to stabilize polarization in $BaFe_{12}O_{19}$, transitioning it from a quantum paraelectric bulk state into a polar state via epitaxial strain[19]. Bulk $BaFe_{12}O_{19}$ has been proposed to exhibit quantum fluctuations of $Fe^{3+}$ ions between two off-center positions within the trigonal bipyramidal (TBP) $FeO_5$ coordination polyhedra, suppressing long-range polarization (Fig. 1a)[17,18]. By applying in-plane biaxial compressive strain, it might be possible to stabilize one of the off-center positions of the $Fe^{3+}$ ions, thereby inducing net polarization along the *c*-axis (Fig. 1b).

In support of our experimental strategy, DFT-calculated phonon dispersions of the parent centrosymmetric structure reveal instabilities along two phonon branches with modes dominated by off-center $Fe^{3+}$ displacements within trigonal bipyramids (Methods and Supplementary Fig. 10). These instabilities intensify with increasing compressive strain. Both branches are close in frequency and exhibit minimal dispersion, suggesting weak coupling between local structural distortions.

To explore energetics further, we relax low-symmetry structures containing unstable modes at high-symmetry *k*-points Γ (0,0,0), M (1/2,0,0), and K (1/3,1/3,0) under varying strain conditions. Figure 1c illustrates the computed energy differences for these configurations, referenced to the ferroelectric (↑) state. Throughout a broad strain range, dipoles preferentially align between planes and anti-align within the plane. Notably, several metastable states, particularly those with parallel dipoles between planes, differ by only a few meV per formula unit from the ground state, reflecting a small energy penalty for dipole flipping into a ferroelectric state. This feature, though atypical in traditional (anti-)ferroelectrics, aligns with the flat phonon branches identified (Supplementary Fig. 10). When compressive strain exceeds 5%, the ferroelectric state (↑) emerges as energetically favorable, consistent with the computational results by Wang *et al*[19].

To comprehensively map the complex strain-dependent energy landscape, we developed an effective lattice Hamiltonian incorporating both long-range dipole-dipole and short-range interactions (Methods and Supporting Information). Fitted against DFT energies (squares in Fig.

1c), our Hamiltonian reliably captures the ground state and reproduces the subtle energy differences observed (shown as the curves in Fig. 1c). Along the out-of-plane direction, both dipole-dipole and short-range interactions favor dipole alignment. Within the plane, however, dipole-dipole interactions instead favor anti-alignment. At lower strain levels, this in-plane dipole-dipole repulsion slightly outweighs the short-range alignment tendency, leading to predominantly anti-aligned dipoles. The near balance of these two competing interactions accounts for the abundance of low-lying metastable states. As strain increases, short-range interactions become dominant, overcoming the dipole-dipole repulsion and stabilizing aligned dipoles within the plane, thus establishing a ferroelectric ground state when the compressive strain exceeds 5%.

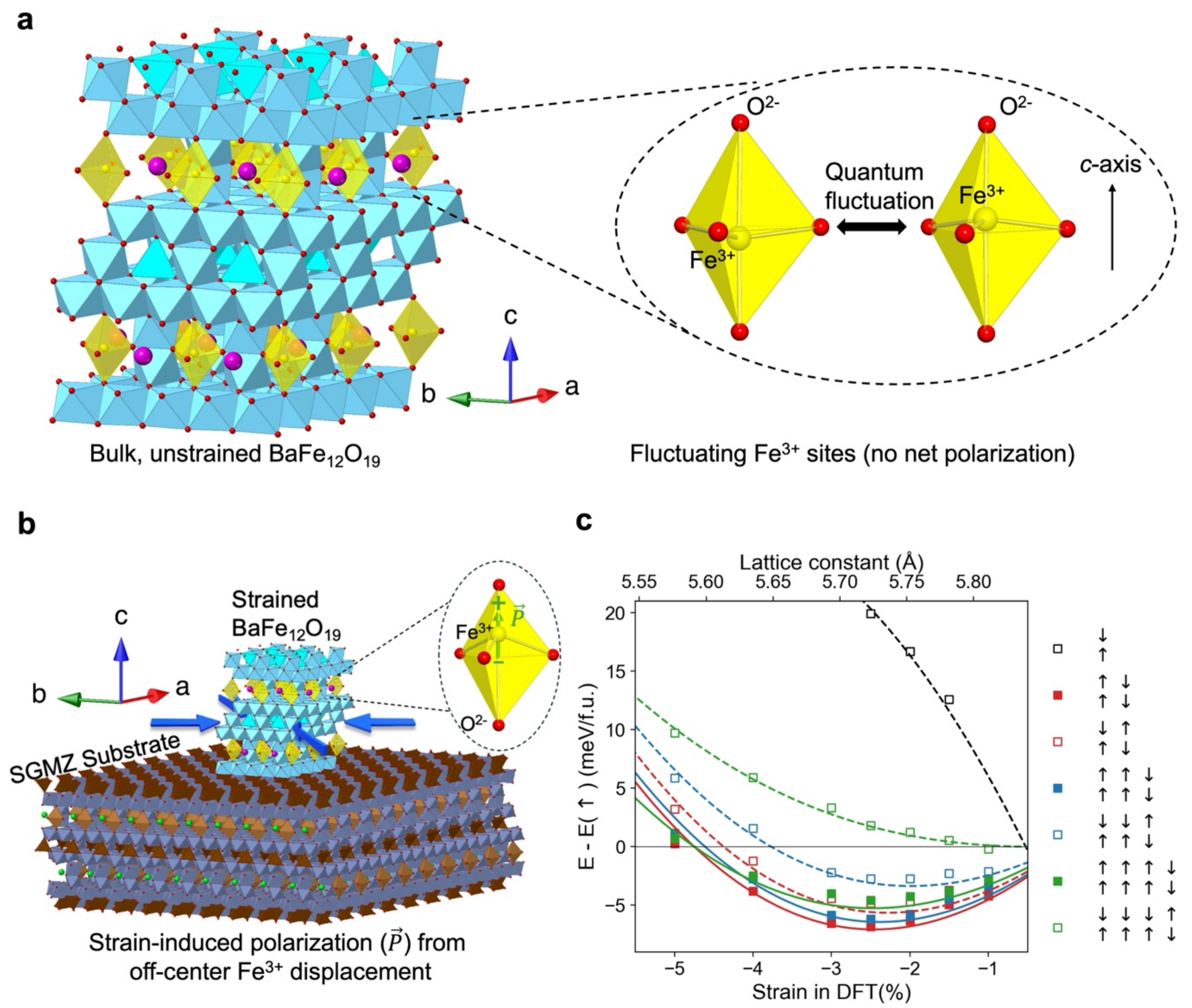

**Fig. 1 | Strain-induced polarization in $BaFe_{12}O_{19}$: Structural origin and DFT energetics.** **a,** Crystal structure of bulk, unstrained $BaFe_{12}O_{19}$ showing trigonal bipyramidal $FeO_5$ units (colored in yellow) with $Fe^{3+}$ ions fluctuating between two off-center positions along the *c*-axis. This quantum fluctuation prevents the emergence of long-range polarization, resulting in a quantum paraelectric state with no net polarization. **b,** Under epitaxial in-plane, biaxial, compressive strain from an SGMZ substrate, $Fe^{3+}$ ions in the $FeO_5$ bipyramids are stabilized in one off-center position, leading to a net polarization $\vec{P}$ along the *c*-axis. This strain-induced symmetry breaking transmutes $BaFe_{12}O_{19}$ into a polar state. **c,** DFT total energies of several electric dipole configurations with reference to ferroelectric (↑) configuration for compressive biaxial strain from 1% to 5%, shown as squares. The variations in DFT total energies are well

captured with a strain-dependent effective lattice Hamiltonian. The energies given by this model Hamiltonian are shown by the curves. The arrows in the legend represent the nearest-neighbor dipole arrangements. For instance, $\begin{smallmatrix}\downarrow & \downarrow & \uparrow \\ \uparrow & \uparrow & \downarrow\end{smallmatrix}$ means the electric dipoles are ↑↑↓ (↓↓↑) in the bottom (top) layer of a triple-sized unit cell, corresponding to a distortion at the K(1/3,1/3,0) point. For clear visualization, we distinguish configurations in which dipoles are arranged parallel and anti-parallel out of plane with solid and hollow square symbols, respectively.

## 3 Growth of Commensurately Strained (001) $BaFe_{12}O_{19}$ Films on (001) $Sr_{1.03}Ga_{10.81}Mg_{0.58}Zr_{0.58}O_{19}$ (SGMZ) Substrate

$BaFe_{12}O_{19}$ thin films were grown by reactive-oxide molecular-beam epitaxy (MBE) on a recently perfected large-diameter SGMZ substrate, which is isostructural to $BaFe_{12}O_{19}$ and exhibits a 1.1% smaller in-plane lattice constant[21]. Precise stoichiometry control during film growth is challenging but essential for achieving a commensurately strained film. A method utilizing *in situ* reflection high-energy electron diffraction (RHEED), X-ray diffraction (XRD), and atomic force microscopy (AFM) to determine the optimal iron-to-barium flux ratio is described in Methods. Using this method, four films in which the majority of the $BaFe_{12}O_{19}$ is commensurately strained were successfully grown on (001) SGMZ substrates. These films range from 22.5 nm to 27.5 nm in thickness. X-ray diffraction reciprocal space maps (XRD RSMs) of the films are shown in Supplementary Fig. 1, together with a thicker $BaFe_{12}O_{19}$ film that is clearly fully relaxed.

The characterization of the 27.5 nm thick (001) $BaFe_{12}O_{19}$ film is presented in Fig. 2. Figure 2a presents a high-angle annular dark-field scanning transmission electron microscopy (HAADF-STEM) image of the interface between the (001) $BaFe_{12}O_{19}$ film and the (001) SGMZ substrate. The atomic columns in the $BaFe_{12}O_{19}$ film are clearly aligned with those of the underlying SGMZ substrate, confirming the predominant coherent, commensurate strain in the film. Figure 2b shows an XRD RSM, illustrating that the $11\underline{18}$ reflection peak from the $BaFe_{12}O_{19}$ film shares the same in-plane reciprocal lattice vector component ($Q_x$) with the $11\underline{18}$ peak from the SGMZ substrate. This further confirms that most of the 27.5 nm-thick $BaFe_{12}O_{19}$ film is commensurately strained to the underlying SGMZ substrate. Additional characterization by *in situ* RHEED, *ex situ* XRD including a rocking curve of the $BaFe_{12}O_{19}$ film exhibiting a full width at half maximum of 0.007°, and AFM measurements (Supplementary Fig. 2) plus STEM (Supplementary Fig. 3), further verify the high crystalline quality, phase purity, and smooth surface morphology of the $BaFe_{12}O_{19}$ films. The lattice constants of this mainly commensurately strained $BaFe_{12}O_{19}$ film, determined by XRD RSM and STEM, are $a$ = 5.82 Å and $c$ = 23.42 Å (see Supplementary Fig. 4 for details).

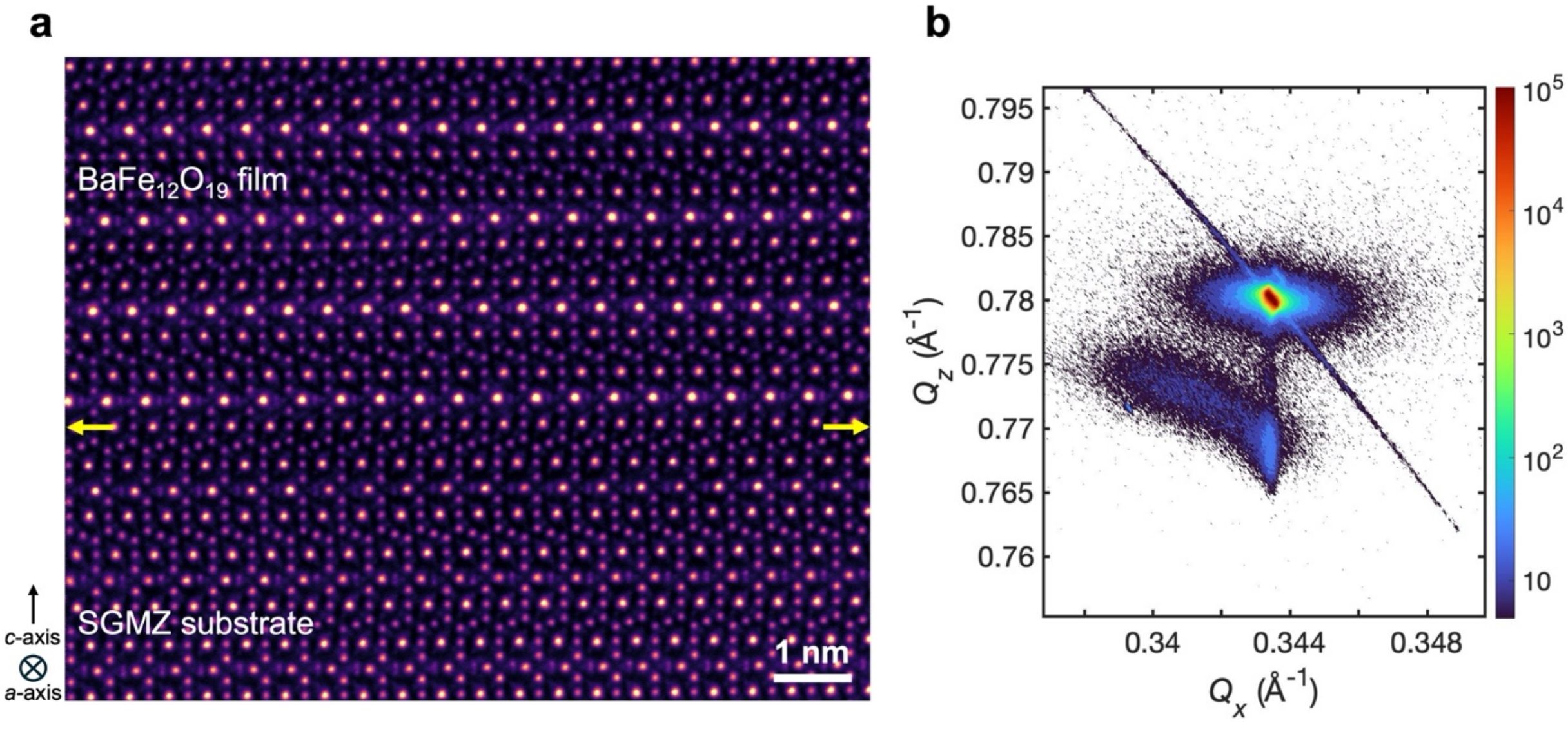

**Fig. 2 | Epitaxial and predominantly commensurate 27.5 nm thick $BaFe_{12}O_{19}$ film on SGMZ.** **a,** High-angle annular dark-field scanning transmission electron microscopy (HAADF-STEM) image showing the sharp and coherent interface between the (001) $BaFe_{12}O_{19}$ film and the (001) SGMZ substrate. **b,** X-ray diffraction reciprocal space map (RSM) around the $11\underline{18}$ reflection,

showing that the $BaFe_{12}O_{19}$ film and SGMZ substrate share the same in-plane reciprocal lattice vector component ($Q_x$), confirming that the film is mainly commensurately strained to the substrate.

## 4 Symmetry Breaking by Optical Second Harmonic Generation

To reveal the symmetry breaking in the nominally commensurately strained $BaFe_{12}O_{19}$ films, temperature-dependent second harmonic generation (SHG) measurements were carried out. Electric dipole SHG is a nonlinear optical process in which incident light at a fundamental frequency $\omega$ is converted into light at twice the frequency, $2\omega$, as a result of broken inversion symmetry in the material, making it a powerful probe of ferroelectric and polar order[22]. A schematic of the setup used for SHG measurements is shown in Fig. 3a. By rotating the incident polarization of light at the fundamental frequency, SHG polar plots can be measured corresponding to two orthogonally polarized (*s*- and *p*-polarized) second harmonic light rays reflected from the sample surface; this is called SHG polarimetry. By fitting the polar plots, the point-group symmetry of materials can be determined. Figure 3b shows the oblique incidence SHG polarimetry for a fully strained $BaFe_{12}O_{19}$ film (biaxial strain, $\varepsilon \approx -1.1\%$) at 300 K, fitted to a $6mm$ point group model, validating that the strain breaks inversion symmetry and stabilizes a polar phase of $BaFe_{12}O_{19}$ at room temperature. The unstrained parent phase has a centrosymmetric $\frac{6}{m}mm$ point group and thus does not generate SHG signal. SHG measurements on a single crystal of $BaFe_{12}O_{19}$ are presented in Supplementary Fig. 6, where no SHG signal was detected. The $6mm$ symmetry corresponds to the two DFT-predicted polar states, indicated by the solid green and solid blue squares in Fig. 1c.

To identify polar phase transitions in the strained $BaFe_{12}O_{19}$ thin films, temperature-dependent SHG measurements were performed during heating from 8 K to 1000 K. While the as-grown sample exhibits 6*mm* symmetry at 300 K (Fig. 3b,c), upon cooling to 8 K, the SHG polarimetry indicates symmetry lowering as shown in Fig. 3c. Additionally, the SHG intensity increases, reaching 2 × the room-temperature intensity at 8 K. (Fig. 3d). While a multidomain orthorhombic *mm*2 symmetry provides a good fit to the low-temperature polar plots, this assignment cannot be definitively confirmed, since further lower symmetry models like a multidomain *m* model can also fit the SHG polarimetry. DFT calculations predict that the ground state at this strain level is an orthorhombic antiferroelectric phase with *mmm* point group symmetry (labeled by solid red squares in Fig. 1c, with its schematic crystal structure shown in Supplementary Fig. 7). If a net out-of-plane polarization emerges, the symmetry would be reduced to *mm*2. Thus, at low temperature, the film hosts a polar orthorhombic phase derived from an antiferroelectric ground state. As the sample is heated back to room temperature, the low-symmetry phase persists, and the 6*mm* symmetry is not restored until 400 K (Fig. 3c), exhibiting thermal hysteresis. Interestingly, the strain-stabilized polar 6*mm* phase remains stable upon further heating up to 1000 K (Fig. 3c), with no appreciable change in SHG intensity above room temperature (Fig. 3d). This signifies the robustness of the strain-stabilized polar phase upon heating, with the phase transition temperature pushed beyond 1000 K under a modest compressive strain of approximately -1.1%.

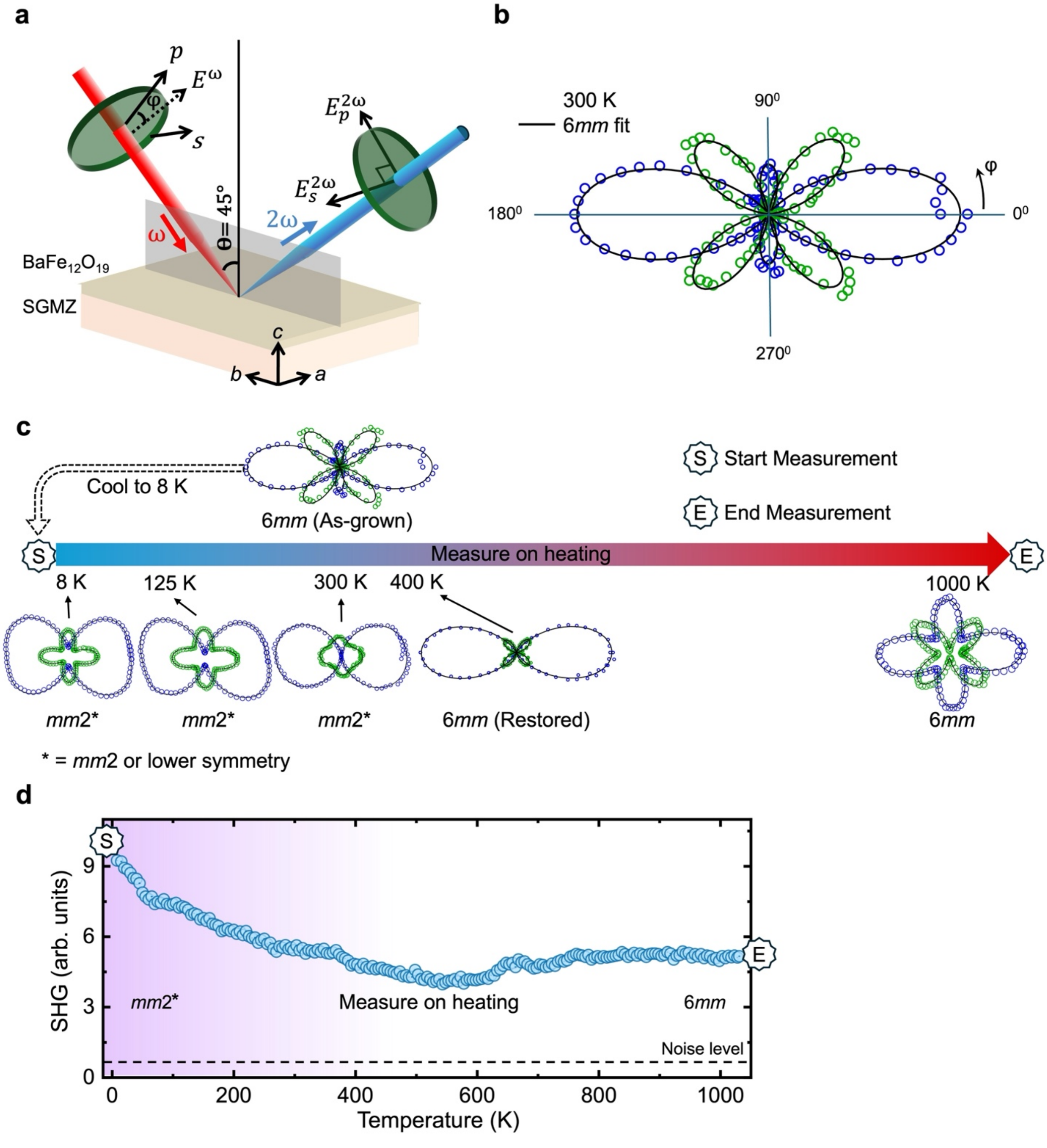

**Fig. 3 | Strain-Stabilized polar phase revealed by temperature-dependent SHG measurements. a,** Schematic of second harmonic generation (SHG) setup in reflection geometry. **b,** SHG polar plot for $BaFe_{12}O_{19}$ on SGMZ at 300 K fitted to the 6*mm* model, showing symmetry lowering from the centrosymmetric unstrained parent phase, consistent with theoretical predictions. Equations for SHG modeling are discussed in the Supporting information. **c,** Temperature-

dependent evolution of SHG polarimetry, exhibiting thermal hysteresis. **d,** SHG intensity vs. temperature during heating from 8 K to 1000 K.

## 5 Polar Nanoregions by Multislice Electron Ptychography

We further probe the biaxial-strain-induced changes in local symmetry by directly imaging and comparing the atomic occupation at the trigonal bipyramid sites in single crystal $BaFe_{12}O_{19}$ and the strained film. This atomic-scale characterization presents two challenges: i) previously reported values for the $Fe^{3+}$ off-centering is ~ 0.2 Å, necessitating very high spatial resolution and ii) spatial variations in the occupation need to be resolved in 3D so that we are sensitive to local symmetries rather than a projected image that gives a spatial average. We address both these challenges using multislice electron ptychography (MEP)[23,24], with a lateral resolution enabling us to resolve the off-centering of the $Fe^{3+}$ ions and a depth resolution of ~2 nm. While this depth resolution is insufficient to resolve the TBP occupation at every unit cell in 3D, we are still sensitive to nanometer-scale polar regions as we demonstrate below.

Figure 4a shows a comparison of projected MEP images of the bulk unstrained single crystal and the strained thin film, where the TBP sites are labelled with a blue outline. In both images, the TBP sites appear elongated suggesting atomic occupancy on both off-center sites. Larger field of view images are provided in Supplementary Fig. 8. A small region cropped from each image in Fig. 4a examining the TBP sites in more detail is shown in Fig. 4b. While the two sites are barely resolved in the single crystal, they appear well-separated in the strained film, indicating a strain-induced increase in the magnitude of off-centering at the TBP sites. This projected view still does not tell us whether the observed off-centering arises from quantum fluctuations or is instead a static distortion. We utilize the 3D imaging capability of MEP to answer this question by analyzing the changes in TBP site occupation along the depth direction as illustrated in Fig. 4c. MEP produces a series of images/slices that show how the sample structure changes along the depth direction; four such images cropped around a TBP site from different depths as shown in Fig. 4c are presented in Fig. 4e. The depth profile along the apical oxygen—TBP iron—apical oxygen line is shown in Fig. 4d and reveals the changing occupation at the TBP site along the depth direction. The $Fe^{3+}$ ion is predominantly in the lower (upper) site at Depth 1 (Depth 4), indicating the presence of local polar regions with ↓(↑) polarization at these depths. At Depth 3, both sites appear occupied, indicating that the polarity changes at a length-scale smaller than the depth resolution of the technique around this depth. At Depth 2, the occupation is peaked between the two sites on the mirror plane, indicating that the off-centering might be suppressed in some unit cells.

In Fig. 4f, we plot the average occupancy at the TBP site in red obtained by averaging the line profiles across different off-center sites, with the mirror plane labelled by the black dashed line. The average occupancy shows a bimodal distribution for the thin film, with peaks corresponding to the two off-center positions. Even though we detect nano-polar regions in the single crystal, the average profile shows a single peak, likely from a combination of occupancy on the mirror-plane site and a smaller magnitude of off-centering (that results in a smaller dip in the center of the intensity profile). By fitting a sum of two Gaussians to the average line profile (Supplementary Fig. 9), we calculate a spacing of 0.42 Å and 0.51 Å for the single crystal and thin film, respectively. The epitaxial strain enhances the magnitude of off-centering by 21%, stabilizes polar nanoregions, but is not enough to stabilize long-range polar order.

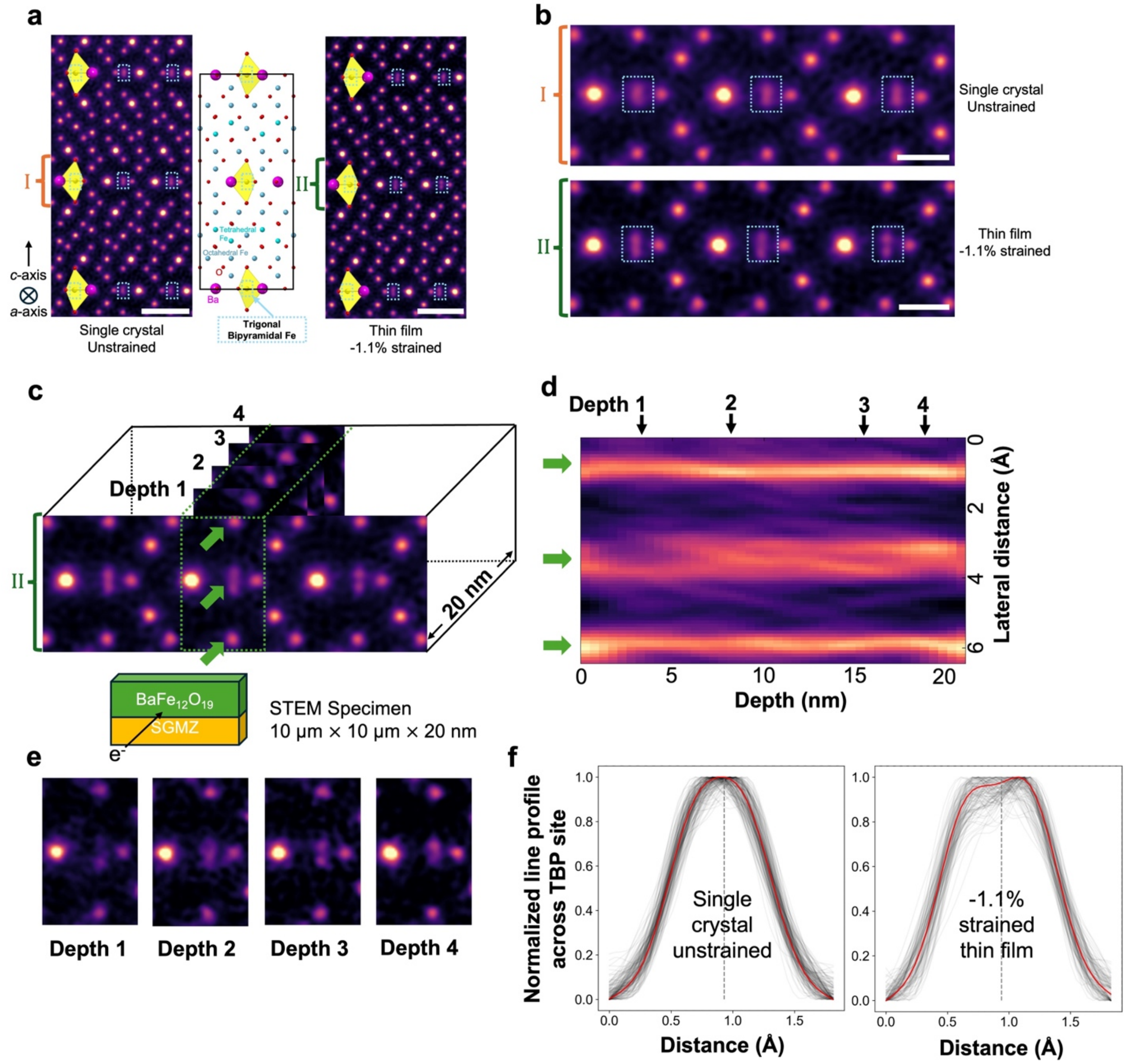

**Fig. 4 | Strain-induced polar nanoregions in $BaFe_{12}O_{19}$ revealed by multislice electron ptychography (MEP). a,** Projected MEP images of the unstrained single crystal (left) and strained thin film (right), with the schematic of one $BaFe_{12}O_{19}$ unit cell in the center. Trigonal bipyramidal (TBP) iron sites are outlined with blue dashed boxes. Scale bars are 5 Å. **b,** Enlarged views of the selected regions (I and II) from panel a. Scale bars are 2 Å. **c,** Schematic of a representative TBP iron site and apical oxygen atoms (marked by green arrows) in the reconstructed object from MEP, showing how the different image slices can be used to track changes in occupancy along the beam direction. Each image corresponds to the structure averaged in the depth direction by around 2 nm (corresponding to the depth resolution of MEP). **d,** Intensity profiles across the apical oxygen–

TBP iron–apical oxygen axis at different depths, extracted along the green arrows in panel c. **e,** Depth slices (1–4) showing site-resolved TBP occupancy at increasing depths. **f,** The average occupancy at the TBP iron site in red obtained by averaging the line profiles across different off-center sites, with the mirror plane labelled by the grey dashed line.

# 6 Corroborating Polar Nanoregions with Path-Integral Monte Carlo Simulations

To further substantiate the existence and dynamics of polar nanoregions revealed by MEP in Section 5, we performed *ab initio* simulations using path-integral Monte Carlo (PIMC) methods based on the effective Hamiltonian (Methods and Supplementary Information). Figure 5 summarizes how the double-well potential profile and the TBP iron distribution evolve with increasing compressive strain at 300 K. As the strain increases, the double-well potential (red curves) becomes deeper and more separated, while the TBP iron distribution exhibits enhanced bimodality with reduced probability density near the center. This trend is in excellent agreement with the increase in off-centering displacement observed by MEP in the strained $BaFe_{12}O_{19}$ film. The full evolution of the TBP iron distribution across temperatures for a range of strain levels is presented in Supplementary Fig. 15. Additionally, Supplementary Fig. 16 demonstrates a clear suppression of quantum fluctuations with increasing compressive strain at low temperatures where quantum fluctuations are more pronounced, consistent with the deepening of the double-well potential. This suppression indicates a reduced tendency for the TBP iron ions to tunnel between off-centered positions, thereby favoring more stable local polarization and the formation of polar nanoregions.

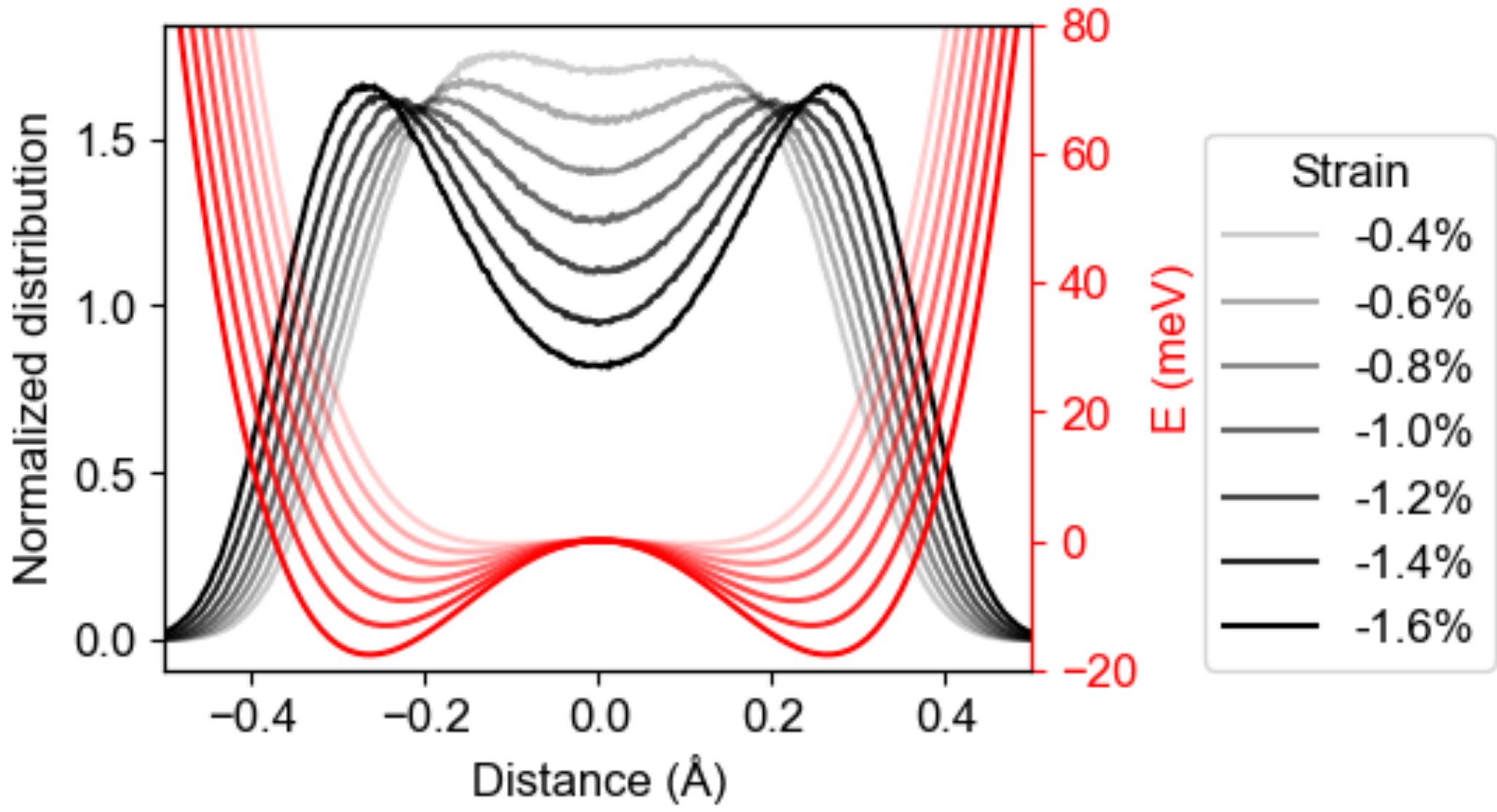

**Fig. 5 |** Normalized distribution of iron in the trigonal bipyramid obtained from PIMC, under various values of compressive strain and at 300 K. The red curves show the double well potential in the ferroelectric configuration.

# 7 Conclusion

In summary, we demonstrate that moderate biaxial compressive strain can induce polar nanoregions in $BaFe_{12}O_{19}$, a ferrimagnetic incipient ferroelectric in its bulk form, thereby enabling room-temperature relaxor multiferroicity in a single-phase material. First-principles calculations reveal a strain-tunable energy landscape with low energy barriers between competing dipolar configurations. Using a structurally matched SGMZ substrate to apply a 1.1% in-plane, biaxial, compressive strain, we achieve predominantly coherent epitaxial growth and confirm symmetry breaking through second harmonic generation and the emergence of polar nanoregions via multislice electron ptychography. While these results collectively indicate the stabilization of local polar order, the applied strain remains below the predicted threshold for achieving long-range ferroelectricity, representing a key limitation of this study. Nevertheless, our work establishes $BaFe_{12}O_{19}$ as a compelling platform for exploring strain-driven multiferroicity and highlights the potential of epitaxial strain engineering to realize new room-temperature multiferroics.

## Acknowledgements

Y.E.L., H.K.P., R.R., D.A.M., and D.G.S. acknowledge support from the Army Research Office under the ETHOS MURI via cooperative agreement W911NF-21-2-0162. M.B., C.G., and D.G.S. acknowledge the Joint Lab between Cornell University and the Leibniz-Institut für Kristallzüchtung that facilitated this research. This research was funded in part by the Gordon and Betty Moore Foundation's EPiQS Initiative through Grants GBMF3850 and GBMF9073. The electron microscopy studies made use of the Cornell Center for Materials Research (CCMR) facilities supported by NSF MRSEC program (DMR-1719875), NSF MIP (DMR-2039380), NSF-MRI-1429155, and NSF (DMR-1539918). The authors also thank John Grazul, Mariena Silvestry Ramos, and Malcolm Thomas for technical support and maintenance of the electron microscopy facilities. S.H. and V.G. Acknowledge support from DOE-BES grant DE-SC0012375 for optical second harmonic generation measurements.

## Methods

**Growth and characterizations of $BaFe_{12}O_{19}$.** Thin films of $BaFe_{12}O_{19}$ were grown by reactive-oxide molecular-beam epitaxy (MBE) in a Veeco GEN10 system. (001) sapphire substrates were annealed in air at 1000 °C for 6 h in a tube furnace and subsequently used as calibration substrates to establish the optimal growth conditions. As-polished (001) hexaferrite substrates $Sr_{1.03}Ga_{10.81}Mg_{0.58}Zr_{0.58}O_{19}$ (SGMZ), were subsequently used for the growth of strained $BaFe_{12}O_{19}$ films. To ensure iron is oxidized into the +3 valent state, distilled ozone (~80% $O_3$ + 20% $O_2$ reaching the substrate) was used as the oxidant and the background pressure of it was kept at $1.0 \times 10^{-6}$ Torr during growth. $BaFe_{12}O_{19}$ was grown by co-deposition of barium and iron at a substrate temperature of 750 °C measured by an optical pyrometer operating at a wavelength of 1550 nm. Precise stoichiometry control is essential for achieving a fully strained film. A method utilizing *in situ* reflection high-energy electron diffraction (RHEED), X-ray diffraction (XRD), and atomic force microscopy (AFM) to determine the optimal iron-to-barium flux ratios is described below. RHEED patterns were recorded using the KSA-400 software and a Staib electron source operated at 13 kV and a filament current of 1.5 A. XRD scans were measured with a PANalytical Empyrean diffractometer with Cu $K\alpha_1$ radiation. AFM surface topography images were taken with the Asylum Research Cypher ES Environmental AFM.

Iron and barium were evaporated from elemental sources at initial fluxes of $3 \times 10^{13}$ and $2.5 \times 10^{12}$ atoms $cm^{-2}$ $s^{-1}$, respectively. Flux calibration methods for iron and barium, with an absolute accuracy of ±1%, were detailed in our previous work[25]. The initial flux ratio of 12:1 did not successfully transfer into a stoichiometric $BaFe_{12}O_{19}$ film, warranting additional calibration based on feedback from *in situ* RHEED, XRD, and AFM. *In situ* RHEED and XRD effectively detected deviations greater than ~5% from stoichiometry. Specifically, half-order streaks appeared in RHEED along the [110] azimuth of the (001) $BaFe_{12}O_{19}$ film under iron-rich conditions (Supplementary Fig. 5a), while an impurity XRD peak at $2\theta \approx 27.6°$ emerged under barium-rich conditions (Supplementary Fig. 5b), which may be associated with the formation of a Ba-rich secondary phase, possibly $BaFe_2O_4$, as suggested by the $BaO–Fe_2O_3$ phase diagram[26]. AFM images further reveal that the iron-rich and barium-rich films exhibit markedly different surface morphologies. The iron-rich film displays scattered mounds with significant height variations, indicative of three-dimensional island growth, whereas the barium-rich film shows a densely packed array of round particles with a smaller height contrast (Supplementary Figs. 5c, 5d). AFM rms roughness served as the key parameter for fine-tuning the iron-to-barium flux ratio when the film was near stoichiometric, as RHEED and XRD no longer indicated the presence of impurities. The AFM rms roughness as a function of the iron-to-barium flux ratio exhibited a valley, with the optimal ratio yielding the smallest rms value, consequently, the smoothest film surface. The $BaFe_{12}O_{19}$ calibration films, 30 nm thick and grown with the optimal flux ratio on (001) sapphire substrates, generally exhibited AFM rms values below 500 pm. Once the optimal iron-to-barium flux ratio was determined, (001) SGMZ substrates were used to grow commensurately strained $BaFe_{12}O_{19}$ films. XRD RSM was performed on the $11\underline{18}$ peaks of the $BaFe_{12}O_{19}$ films and SGMZ substrates to determine whether $BaFe_{12}O_{19}$ is fully strained.

**Scanning Transmission Electron Microscopy (STEM)**

STEM lamellae were prepared using the standard lift-out method on a Thermo Fisher Helios G4 UX focused ion beam tool. All STEM data were acquired on an aberration-corrected Thermo Fisher Spectra X-CFEG STEM with 300 kV accelerating voltage, probe semi-convergence angle of 30 mrad, and beam current of 60 pA. The 4D-STEM data for MEP was acquired on a 128 x 128 pixel EMPAD[27] detector with a frame time of 1 ms, scan step-size of 0.42 Å, probe overfocus of 10 nm, an outer collection angle of 53.3 mrad and 256 x 256 scan steps. MEP reconstructions were done using the maximum-likelihood multislice ptychography algorithm implemented in the fold-slice package[23,28].

**First-principles calculations**. Density functional theory (DFT) calculations were performed using the VASP code[29-32] with projector augmented wave formalism[33] and the PBEsol exchange–correlation functional[34]. A 500 eV kinetic energy cutoff and an $8 \times 8 \times 2$ $k$-mesh were used for the primitive cell. All calculations were spin-polarized with Hubbard $U = 5$ eV and $J = 1$ eV for Fe $3d$ orbitals, and the ferrimagnetic configuration was adopted as observed experimentally[16]. Structures were relaxed until forces on atoms were below $10^{-3}$ eV/Å. For compressively strained structures, only the lattice constant and internal coordinates along the $c$ direction were relaxed.

Phonon dispersions were obtained from $2 \times 2 \times 1$ supercells using *phonopy*[35,36] as shown in Supplementary Fig. 10. We find two branches that become increasingly unstable with greater in-plane compressive strain. Along the branches, the phonon displacements mainly consist of vertical displacement of iron in the trigonal bipyramid, forming electric dipoles. The two branches are relatively flat across the Brillouin zone, suggesting weak interactions between these localized dipoles. Further calculations show that the phonon frequencies are largely affected by out-of-plane lattice expansion, which elongates the trigonal bipyramid along the $c$-axis, rather than the in-plane compression. This is consistent with the vertical displacement pattern of iron.

**Effective lattice Hamiltonian.** An effective lattice Hamiltonian consisting of the local mode, dipole-dipole, and short-range interactions[37] was constructed (Supplementary Table 1). The coefficients are strain dependent, with $\varepsilon$ being percentage of biaxial in-plane strain, i.e., $\varepsilon = -1$ when the $a$ lattice constant is 1% smaller than the bulk (unstrained) value. We use $\vec{u}$ to represent the local polar mode, which is mainly due to the iron displacement along the $c$-axis in the trigonal bipyramid. Therefore, $\vec{u}$ between different sites are either aligned or anti-aligned ($\hat{u}_a \cdot \hat{u}_b = \pm 1$). The magnitude of $\vec{u}$ varies with strain but only differs slightly across different sites in relaxed cells.

Therefore, we assume it to be the same on all sites to simplify the Hamiltonian, i.e., $|u_a| = |u_b| = |u|$ for any site $a, b$. This assumption is further supported by the weak interactions between the local dipoles deduced from the phonon dispersion. The value of A in the strain-energy term $A\varepsilon^2$ with $A = 83.96$ meV that describes the energy variation of the high-symmetry phase with respect to strain but has no $u$ dependence, is treated as a constant.

The local-mode energy is obtained using the ferroelectric configuration. Once the polar mode $u$ is present, there appears a mode that transforms as the identity irrep, call it $\eta$, which further lowers the energy and enhances the polar mode. The energy lowering becomes quite significant at large strain so that $\eta$ cannot be ignored. At each strain, the energy landscape $E(u,\eta)$ can be well described by a Landau expansion to 5$^{th}$ order (Supplementary Fig. 11):

$$E(u,\eta) = C_{20}u^2 + C_{02}\eta^2 + C_{21}u^2\eta + C_{40}u^4 + C_{22}u^2\eta^2 + C_{04}\eta^4 + C_{41}u^4\eta + C_{23}u^2\eta^3. \qquad (1)$$

To eliminate $\eta$, we renormalize the coefficients by finding the minimum at each value of the polar mode amplitude $u$, i.e., numerically solving $\partial E/\partial\eta = 0$, shown by the red curve in Supplementary Fig. 11. For each strain, this curve is then fit with the function $\alpha u^2 + \beta u^4$, summarized in Supplementary Table 2, and we find a linear relation between quadratic coefficient $\alpha$ and strain $v$. The quartic coefficient $\beta$, however, is hardly affected by strain.

The dipole-dipole matrix elements are

$$Q_{ab,\alpha\beta}(\varepsilon) = \frac{\pi}{\Omega(\varepsilon)} \sum_{\vec{G}\neq 0} \frac{1}{|\vec{G}(\varepsilon)|^2} \exp\left(-\frac{|\vec{G}(\varepsilon)|^2}{4\lambda^2}\right) \cos\left(\vec{G}(\varepsilon)\cdot\vec{R}_{ab}(\varepsilon)\right) G_\alpha(\varepsilon)G_\beta(\varepsilon) - \frac{\lambda^3}{3\sqrt{\pi}}\delta_{\alpha\beta}\delta_{ab}, \qquad (2)$$

an Ewald summation over the reciprocal lattice vectors $\vec{G}$ that is equivalent but converges faster than the real-space expression[38,39]. Here, $\alpha, \beta = x, y, z$ and $a, b$ are the site labels. $\Omega(\varepsilon)$ is the volume, which is half the volume of a unit cell since there are two trigonal bipyramids in a unit cell. $Z^*(\varepsilon)$ is the phonon mode effective charge and $\epsilon_\infty(\varepsilon)$ is the high-frequency dielectric constant. Because dipoles only appear along the $z$ direction, we only consider the $Q_{ab,zz}$ component and take the $zz$ component of $\epsilon_\infty(\varepsilon)$. $Z^*(\varepsilon)$ and $\epsilon_\infty(\varepsilon)$ are directly fit to the output values by DFT.

Once we find the coefficients for the local mode and dipole-dipole terms, the final step involves getting $J's$ in the short-range terms. The short-range interaction takes into account 5 nearest neighbors, where $J_1$ and $J_2$ are interactions in the same plane, and $J_3$, $J_4$, and $J_5$ are those in adjacent planes, see Supplementary Fig. 12a. The distance between trigonal bipyramids described by $J_5$, which is the longest one, is about 15 Å. Since $J_1$ is significantly larger than the others and is critical in determining the ground state, we set $J_1$ to be strain dependent.

Because we assume that at each strain, the magnitude of $u$ at every site is the same, the quadratic terms of the effective Hamiltonian can be grouped together:

$$\mathcal{H} = \left[\sum_{ab} \alpha(\varepsilon)\delta_{ab} + \sum_{a\neq b} \frac{2Z^*(\varepsilon)^2}{\epsilon_\infty(\varepsilon)} Q_{ab,zz}(\varepsilon)\hat{u}_a \cdot \hat{u}_b + \frac{1}{2}\sum_{\lambda,<a,b>} J_\lambda(\varepsilon)_{ab}\hat{u}_a \cdot \hat{u}_b\right] |u_a||u_b| + \sum_a \beta(\varepsilon)u_a^4$$

$$\equiv \left[N\alpha(\varepsilon) + \left(\textstyle\sum_{a\neq b} \alpha_{DD}(\varepsilon)\hat{u}_a \cdot \hat{u}_b + \sum_{<a,b>} \alpha_{SR}(\varepsilon)\, \hat{u}_a \cdot \hat{u}_b\right)\right]|u|^2 + N\beta(\varepsilon)|u|^4, \quad (3)$$

where $N$ is the total number of sites.

Using the ferroelectric configuration as the starting point, at energy minimum $|u|_{min} = \sqrt{-\alpha(\varepsilon)/(2\beta)}$, the energy is $E_{min} = -\alpha(\varepsilon)^2/(4\beta)$. If flipping dipoles changes the dipole-dipole

and short-range interaction coefficients by $\Delta\alpha(\varepsilon) \equiv \Delta\alpha_{DD}(\varepsilon) + \Delta\alpha_{SR}(\varepsilon)$, the shift in total energy would be

$$\Delta E(\varepsilon) = -\frac{2\alpha(\varepsilon)\Delta\alpha(\varepsilon)+\Delta\alpha(\varepsilon)^2}{4\beta}. \quad (4)$$

This expression is then fit with the DFT calculated total energy, using the ferroelectric configuration as a reference, to extract $J's$. The fitting yields an rms error in energy of 0.34 meV per formula unit (Supplementary Fig. 12b). The magnitude of $J_i$ coefficients falls with increasing distance, i.e., $|J_1| > |J_2|, |J_3| > |J_4| > |J_5|$, consistent with physical intuition. The robustness and extrapolation capability of our Hamiltonian is also proved by additional test points, illustrated in Supplementary Fig. 13a. The energies of these test cases with the 4 ↑ 2 ↓ dipole configuration in the plane and ferroelectric out of plane match well with the energy given by our effective Hamiltonian.

The fitting results indicate competition between the dipole-dipole interaction, which favors anti-parallel dipoles in the plane, and short-range interaction, which favors parallel dipoles in the plane. Supplementary Fig. 13b illustrates the change in the quadratic coefficients when going from the ferroelectric (↑) to the antiferroelectric ($\begin{smallmatrix}\uparrow & \downarrow \\ \uparrow & \downarrow\end{smallmatrix}$) configuration by flipping half of the dipoles. For dipole-dipole interactions, a negative $\Delta\alpha_{DD}$ means that the antiferroelectric ($\begin{smallmatrix}\uparrow & \downarrow \\ \uparrow & \downarrow\end{smallmatrix}$) configuration has lower energy. For the short-range interaction, a positive $\Delta\alpha_{SR}$ suggest the ferroelectric (↑) configuration is energetically favorable. The total $\Delta\alpha_{DD} + \Delta\alpha_{SR}$ changes sign between –4% and –5% strain, favoring the ferroelectric (↑) configuration at large compressive strain.

**Path-integral Monte Carlo (PIMC).** The evolution of the iron separation in trigonal bipyramid and other experimental observables with strain and temperature can be simulated by PIMC based on the effective Hamiltonian. We use a $12 \times 12 \times 5$ supercell ($12 \times 12 \times 10$ sites) and all local modes ($u$) are initialized to be 0. We first perform $1 \times 10^6$ Monte Carlo sweeps (MCS) to reach equilibrium and then $1 \times 10^5$ MCS for statistical averaging to obtain the physical quantities. The number of time slices $N_s$ at temperature $T$ for path integral follows the recipe of Zhang *et al.*, i.e., $N_s = \lceil 600/T \rceil$[40]. In each MCS, we loop over all sites and use the bisection algorithm to improve the sampling efficiency[41].

As the local mode consists mainly of displacement of iron in the trigonal bipyramid, the normalized distribution of iron is approximately that of the local mode, which is shown at a few different strains and temperatures in Supplementary Fig. 14. The top two panels show that as the larger compressive strain makes the double well potential (red curves) wider and deeper, the iron distribution becomes more bimodal, with less weight near the center. At $T$ = 20 K, when the strain reaches –1.4%, the probability of iron near the center is almost 0, suggesting the suppression of quantum paraelectricity as tunneling between the double well is nearly impossible. The bottom two panels, together with Supplementary Fig. 15 reveal the drastically different temperature-dependence of the distribution with increasing strain. At moderate strain, the iron has sufficient kinetic energy to cross the shallow barrier at the center of the double well even at low temperatures, such that it avoids the high energy walls at the two sides by concentrating more near the center, resulting in a reduced average Fe-Fe separation at lower temperatures. As we increase the strain, however, the higher central barrier makes the central positions less favorable, forcing the distribution to be more spread out at lower temperatures.

Above, we inferred subdued quantum paraelectricity at low temperature from the iron distribution, at large compressive strain. Here, we directly estimate the quantum fluctuation, the origin of quantum paraelectricity, from PIMC simulations, providing more evidence for compressive-strain-induced suppression of quantum paraelectricity. The thermal fluctuation is calculated by $\Delta u_{thermal}^2 = \langle\langle u\rangle_s^2\rangle_{i,t}$ and total fluctuation by $\Delta u_{total}^2 = \langle u^2\rangle_{i,s,t}$, where $s, i, t$ are the time slice of imaginary-time path-integral, lattice site, and Monte Carlo sweep in the PIMC simulation. The quantum fluctuation is then estimated through $\Delta u_{total}^2 - \Delta u_{thermal}^2$. In Supplementary Fig. 16, the estimated quantum fluctuation (red curve) contributes significantly to the total fluctuation at low temperature for strain = –0.6% and –0.8%, while becoming very small at larger compressive strains. Since one of the experimental signatures for quantum paraelectricity is a plateau in dielectric susceptibility at low temperature, we then study the effect of strain on dielectric susceptibility $\chi$. At temperature $T$, $\chi$ is related to the ensemble variance of $\bar{p}$ ($\bar{p} = \langle p\rangle_{s,i} = Z^*(\varepsilon)\langle u\rangle_{s,i}$ is the averaged dipole moments over the sites and the time slices) by $\chi = \frac{N_{site}}{\epsilon_0 k_B T V}\left(\langle\bar{p}^2\rangle_t - {\langle\bar{p}\rangle_t}^2\right)$, for $N_{site}$ in the simulation supercell, and each site with volume $V$. To mitigate noise in the result, we average over 40 PIMC simulations at each temperature. The results are plotted in Supplementary Fig. 17. Although the absolute value of $\chi$ is smaller than the experimental value[42], reaching about 40 as the temperature approaches 0 K, the qualitative feature of a plateau is clearly visible for moderate strain. At –1.2% strain and beyond, the plateau disappears, and a peak starts to form, signaling the onset of a phase transition.

We now examine the state below the phase transition temperature, with a real-space dipole correlation map to visualize the relation between nearest neighbors, which is shown for –1.4% strain and at 4 K in Supplementary Fig. 18. The in-plane dipole correlation, $S(\Delta x, \Delta y)$, on each layer of $12 \times 12$ sites is computed through $S(\Delta x, \Delta y) = \sum_{x,y}\langle\bar{u}(x,y)\bar{u}(x+\Delta x, y+\Delta y)\rangle$ and then averaged over all 10 layers and MCS. Here $\bar{u} = \langle u \rangle_s$ and $S(\Delta x, \Delta y)$ is normalized so that $S(0,0) = 1$.

Supplementary Fig. 18 shows some negative correlation between nearest neighbors (–0.299), which is expected for an antiferroelectric ground state, but long-range ordering is still missing due to the geometric frustration on a triangular lattice, as the magnitude of correlation is much smaller than 1. There is a weak positive correlation between next-nearest neighbors (+0.155), an even weaker positive correlation between the third-nearest neighbors (+0.037), but negligible correlation beyond that.

**Second harmonic generation (SHG) measurements**. SHG polarimetry was measured at $\lambda$ = 800 nm fundamental light from a regeneratively amplified Spectra-Physics Solstice Ace Ti:Sapphire laser system (1 kHz, 100 fs). A schematic of the setup is shown in Fig. 3a. Linearly polarized light incident on the sample at an incidence angle $\theta$ = 45° generated second harmonic light at $\lambda$ = 400 nm. The *p*-polarized and *s*-polarized SHG intensities were spectrally filtered and measured by a photomultiplier tube through lock-in amplifier (SR830) detection. Polar plots were generated by rotating the polarization angle ($\varphi$) of the incident fundamental light by a half-wave plate. Temperature-dependent measurements were done through a Janis 300 gas flow cryostat for low-temperature and a home-built heating stage for high-temperature experiments.

# Supporting Information

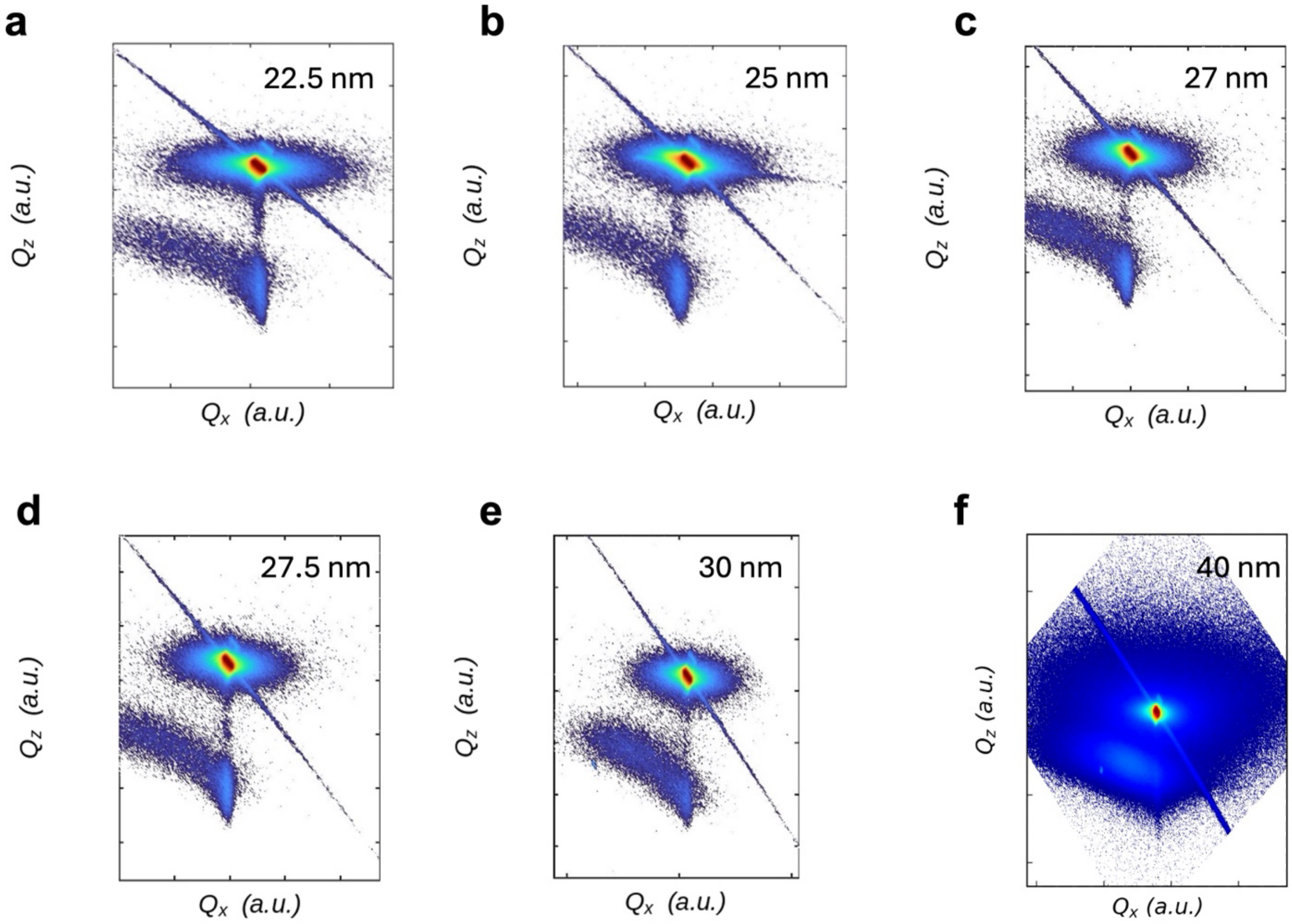

**Supplementary Fig. 1 | Reciprocal space maps of (001) $BaFe_{12}O_{19}$ films of varying thickness on (001) SGMZ.** Asymmetric 11$\underline{18}$ X-ray diffraction reciprocal space maps for $BaFe_{12}O_{19}$ films of thickness (**a**) 22.5 nm, (**b**) 25 nm, (**c**) 27 nm, (**d**) 27.5 nm, (**e**) 30 nm, and (**f**) 40 nm, all grown on (001) SGMZ substrates. In each panel, the intense peak corresponds to the 11$\underline{18}$ reflection of the SGMZ substrate, and the weaker feature corresponds to the $BaFe_{12}O_{19}$ film reflection. For a fully strained (commensurate) film, the film and substrate peaks are aligned vertically along $Q_x$, indicating that the film shares the substrate's in-plane lattice parameter. For films up to 27.5 nm thick (**a–d**), the majority of the film signal remains vertically aligned with the substrate peak, indicating that most of the $BaFe_{12}O_{19}$ is commensurately strained. At 30 nm (**e**), a larger fraction of the film signal has shifted away from vertical alignment, showing that a greater portion of the film has begun to relax. By 40 nm (**f**), the entire diffuse film signal is shifted to smaller $Q_x$ than

the substrate peak, corresponding to a larger in-plane lattice parameter than the substrate and indicating that the film is fully relaxed. $Q_x$ and $Q_z$ axes are in arbitrary units.

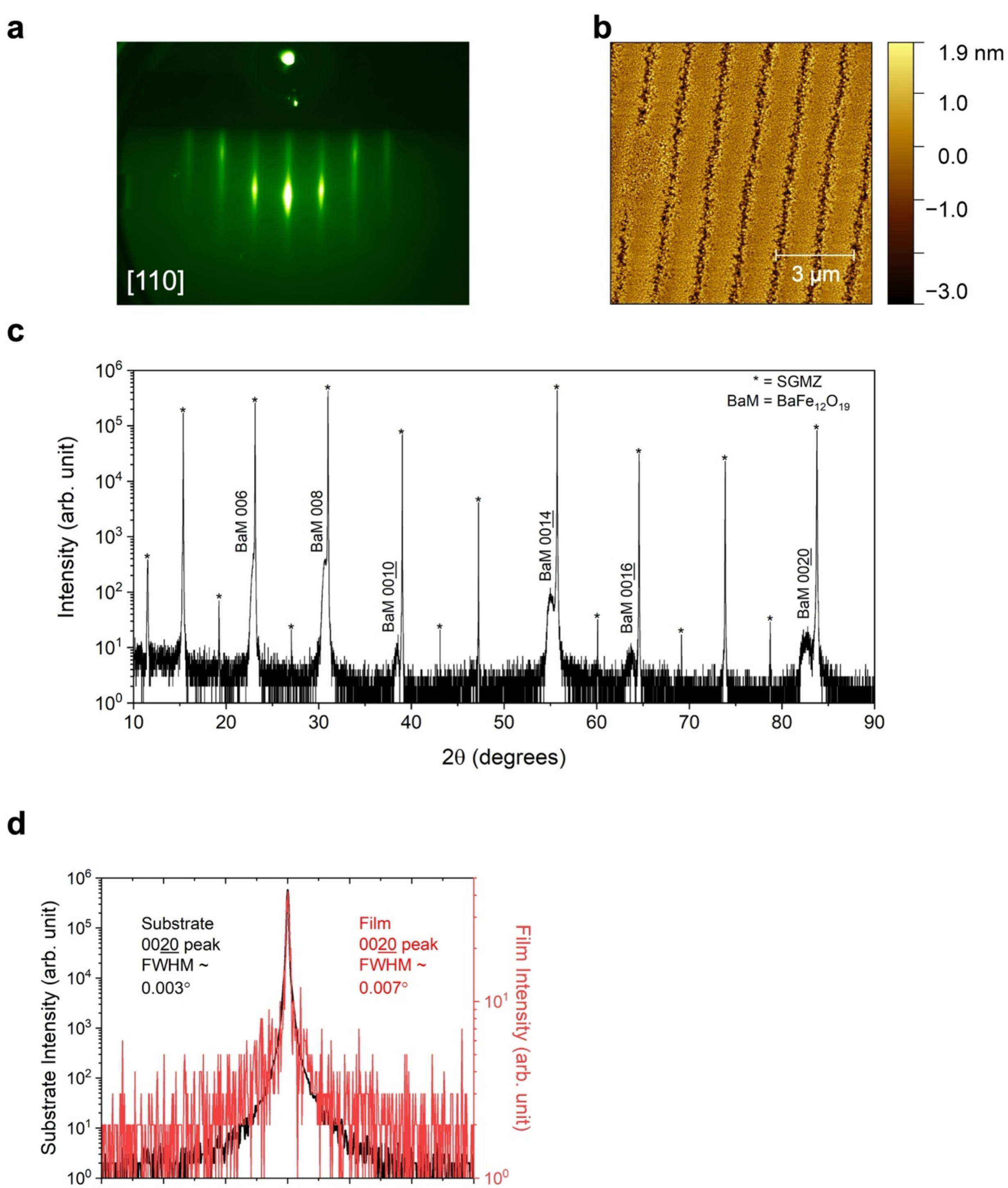

**Supplementary Fig. 2 | Characterization of the 27.5 nm thick (001) $BaFe_{12}O_{19}$ film on a (001) SGMZ substrate.**

**a,** *In situ* RHEED image taken at the end of the growth. **b,** AFM image showing the surface topography, with an rms roughness of 724 pm. **c,** XRD $\theta$-$2\theta$ scan of the film. The asterisk (*) denotes the 00*n* XRD peaks from the (001) SGMZ substrate. **d,** Rocking-curve XRD scan of the 00$\underline{20}$ peak of the $BaFe_{12}O_{19}$ film (red) compared with the 00$\underline{20}$ peak of the SGMZ substrate (black). BaM: $BaFe_{12}O_{19}$. FWHM: full width at half-maximum.

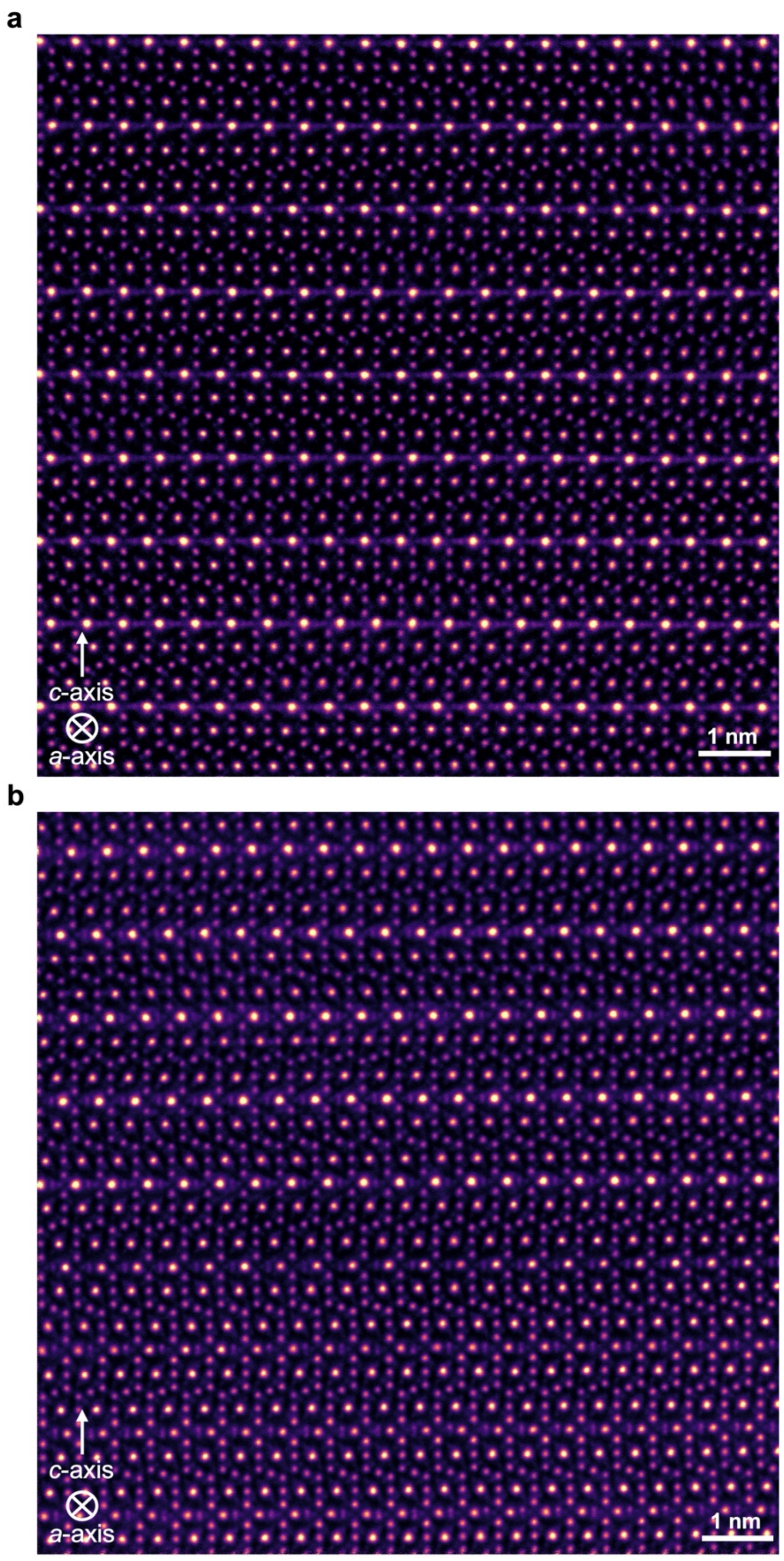

**Supplementary Fig. 3 | HAADF images of single-crystal $BaFe_{12}O_{19}$ (top) and the 27.5 nm thick $BaFe_{12}O_{19}$/SGMZ (bottom).**

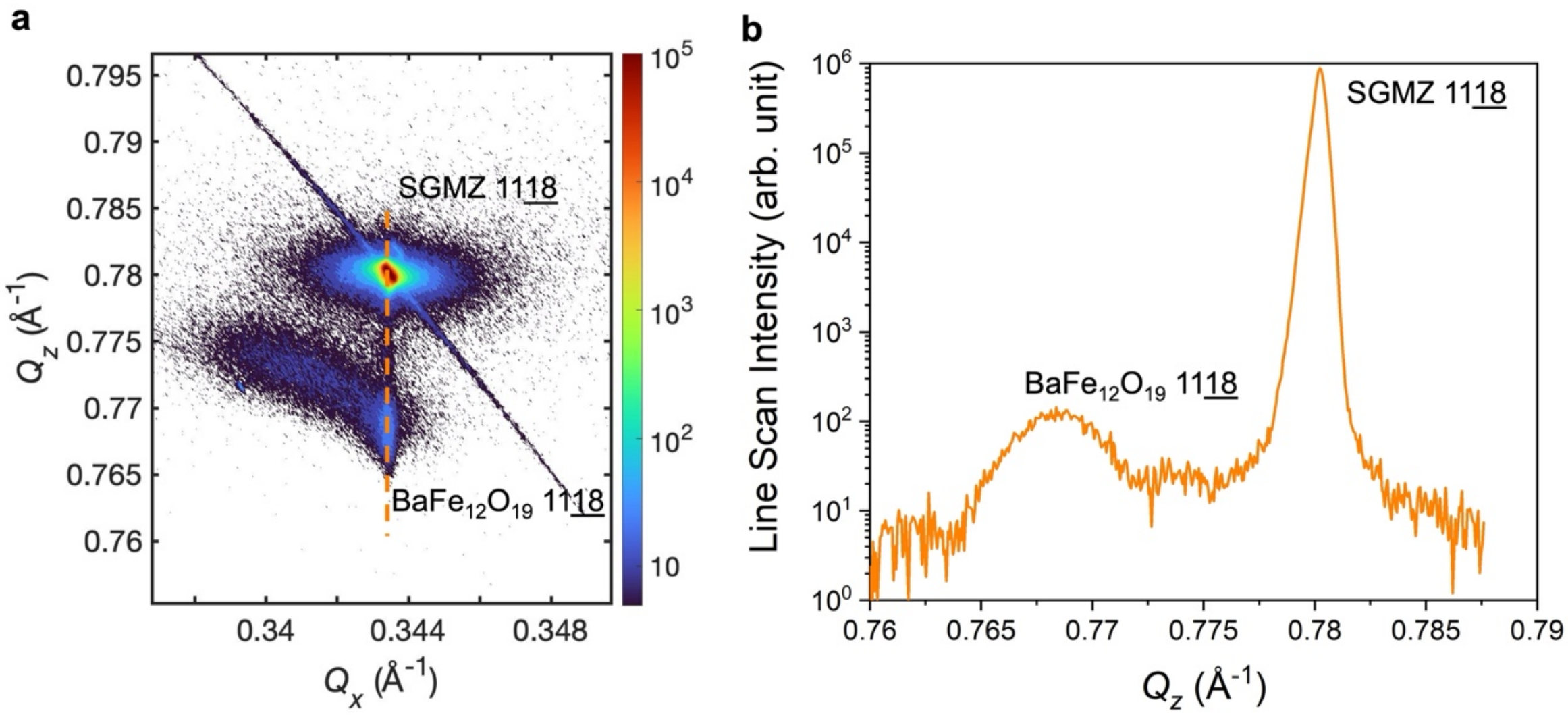

**Supplementary Fig. 4 | X-ray diffraction reciprocal space map (RSM) of the 27.5 nm thick $BaFe_{12}O_{19}$ film grown on (001) SGMZ and a line scan along $Q_z$.**

**a,** RSM around the $11\underline{18}$ reflection of both the $BaFe_{12}O_{19}$ film and the SGMZ substrate. **b,** Line scan along the $Q_z$ direction (orange dashed line in **a**) showing distinct film and substrate peaks. The lattice constants of the $BaFe_{12}O_{19}$ film are calculated to be $a$ = 5.82 Å and $c$ = 23.42 Å, using the lattice constants of SGMZ ($a$ = 5.82 Å and $c$ = 23.07 Å) as a reference.

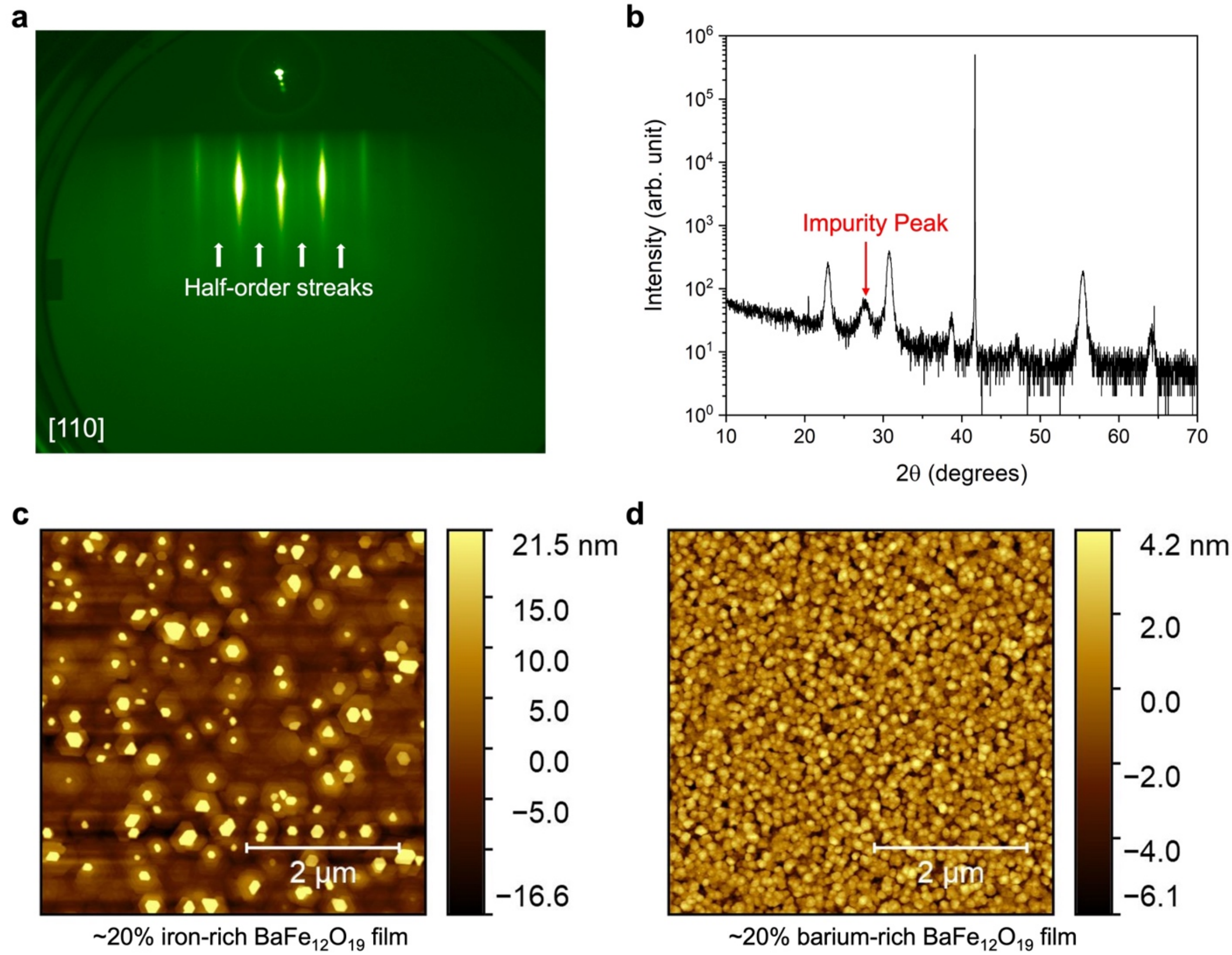

**Supplementary Fig. 5 | Distinctive features of iron-rich and barium-rich $BaFe_{12}O_{19}$ films revealed by RHEED, AFM, and XRD. a,** RHEED pattern along the [110] azimuth of the (001) $BaFe_{12}O_{19}$ film grown under iron-rich conditions (~20% iron rich). The characteristic half-order streaks are indicated by the white arrows. **b,** XRD pattern highlighting an impurity peak at $2\theta \approx 27.6°$ in a $BaFe_{12}O_{19}$ film grown under barium-rich conditions (~20% Ba excess). **c,** AFM image of the ~20% iron-rich film, showing scattered mounds across the surface. **d,** AFM image of the ~20% Ba-rich film, showing densely packed round particles across the surface.

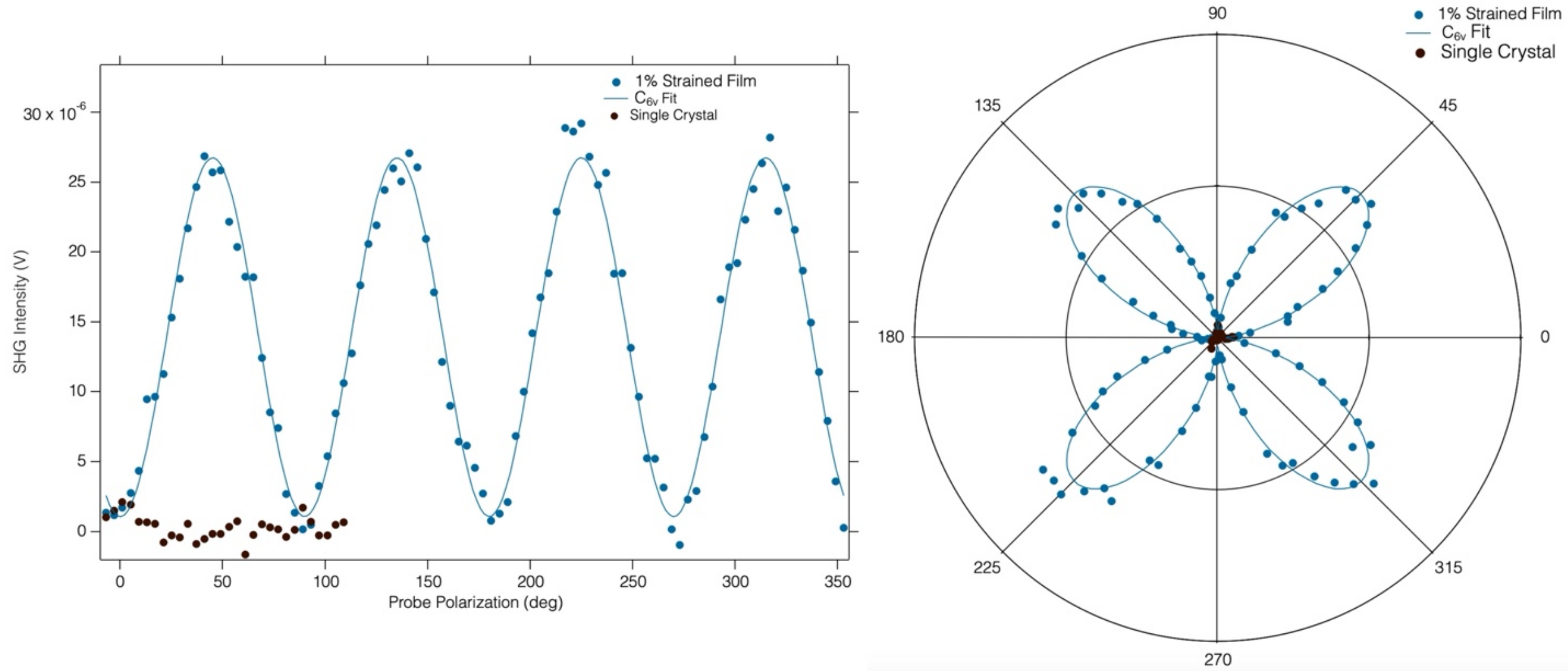

**Supplementary Fig. 6 | SHG polarimetry of a single-crystal $BaFe_{12}O_{19}$ sample and the 27.5 nm thick strained $BaFe_{12}O_{19}$ film.** The strained film exhibits an SHG polarimetry patterns that fit well to the 6*mm* point group symmetry, whereas the single-crystal sample shows no detectable SHG signal due to its centrosymmetric crystal structure. The same probe fluence of approximately 2 mJ/cm$^2$ was used for both the thin film and single crystal samples.

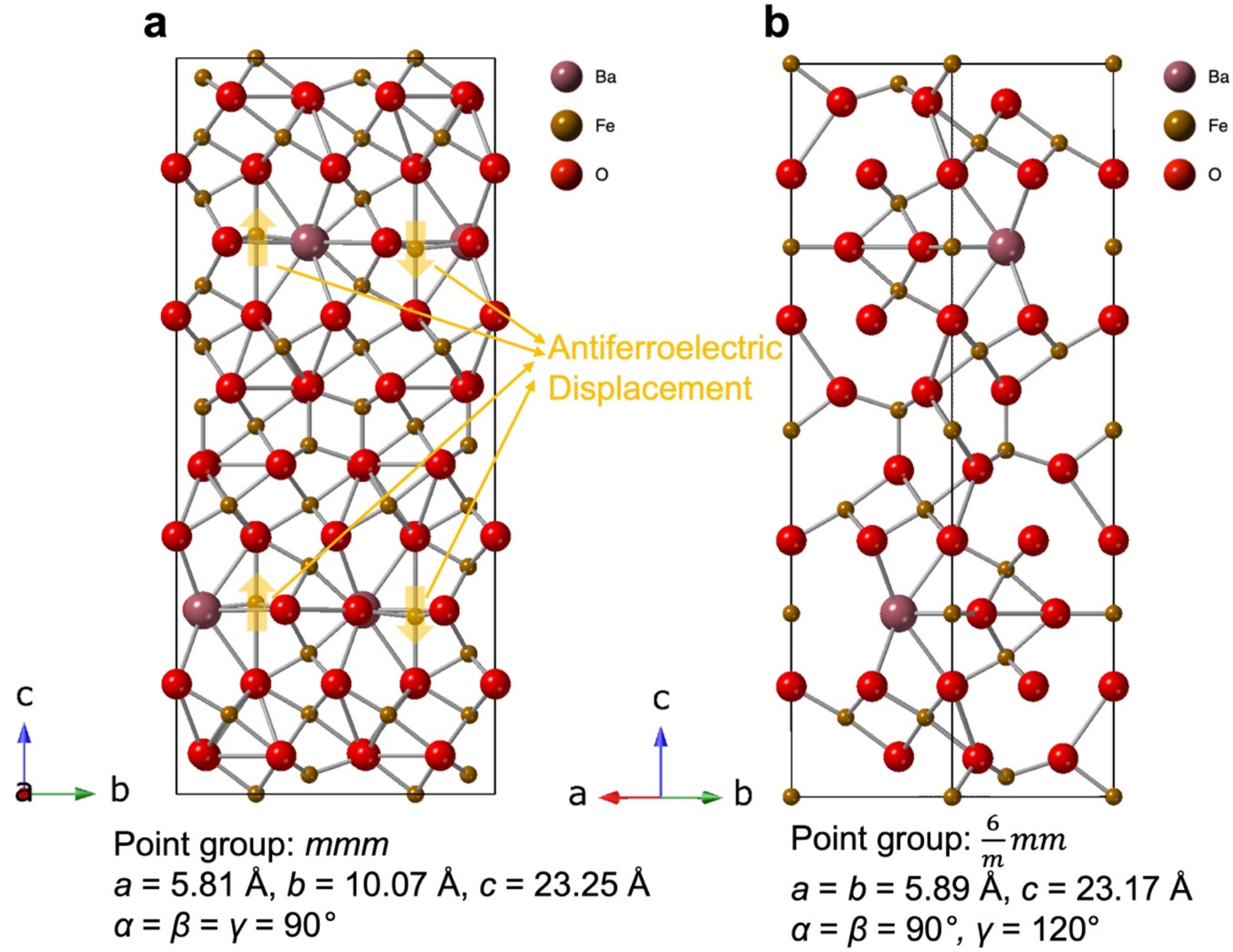

**Supplementary Fig. 7 | Schematic crystal structures of a,** the DFT-predicted ground state at -1% biaxial compressive strain level, an orthorhombic antiferroelectric phase with *mmm* point group symmetry. Semi-transparent yellow arrows mark the antiferroelectric displacement of iron atoms. **b,** the unstrained bulk $BaFe_{12}O_{19}$ phase with a centrosymmetric $\frac{6}{m}mm$ point group.

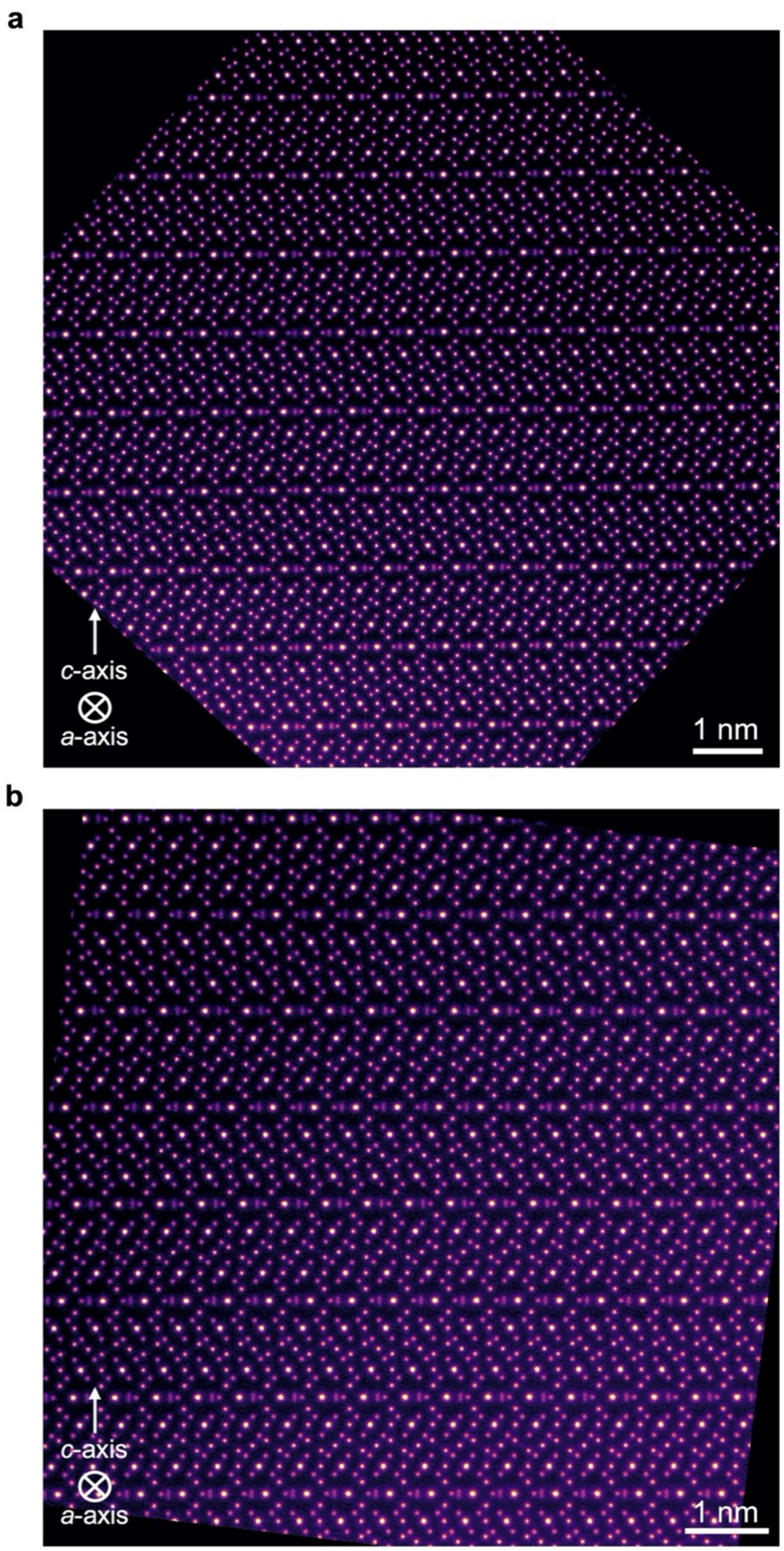

**Supplementary Fig. 8 | MEP images of single-crystal $BaFe_{12}O_{19}$ (top) and the 27.5 nm thick $BaFe_{12}O_{19}$ (bottom).**

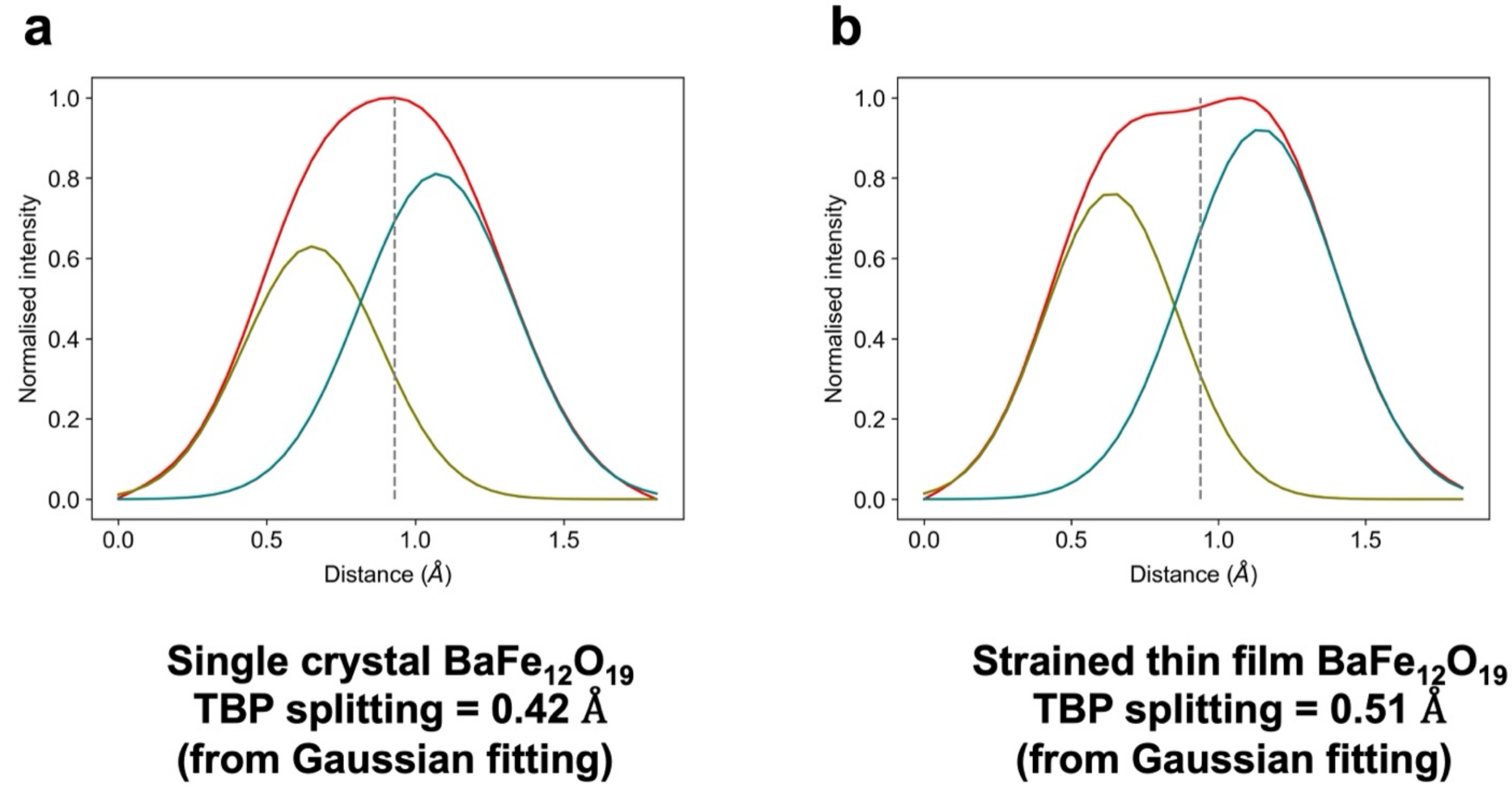

**Supplementary Fig. 9 | Fitting a sum of two Gaussians to the average line profile (Fig. 4f) to calculate the spacing between the two TBP Fe sites for strained and unstrained $BaFe_{12}O_{19}$.**

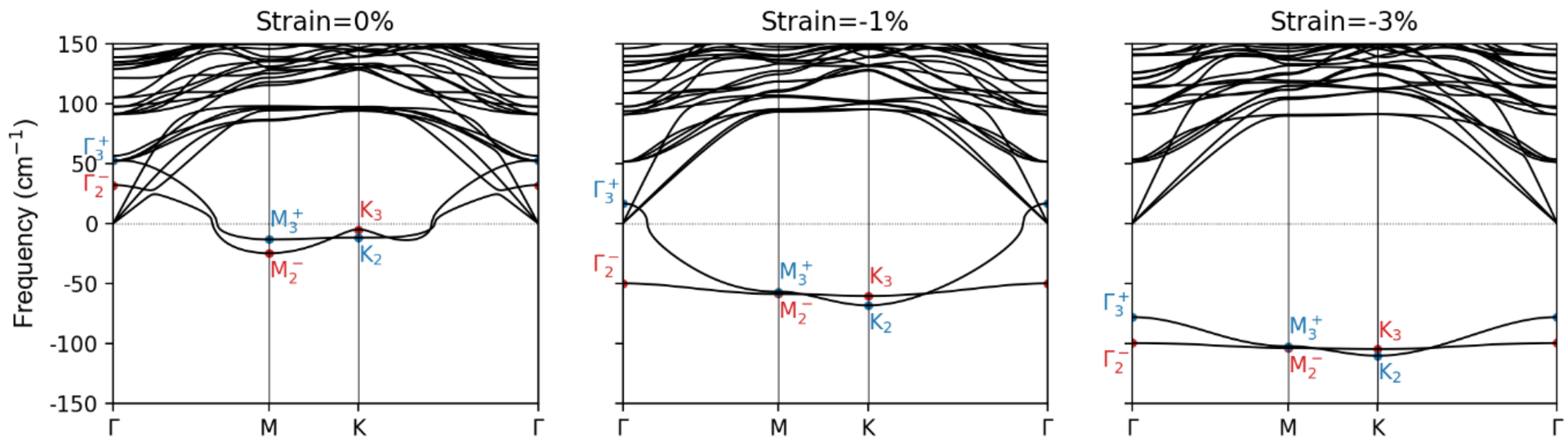

**Supplementary Fig. 10 |** Phonon dispersion of $BaFe_{12}O_{19}$ with compressive strain = 0%, 1%, and 3%. The irreducible representations at high-symmetry points on the two soft branches are labeled.

| Component | Expression | Coefficient | Unit |
| --- | --- | --- | --- |
| Local mode | $\sum_a \alpha(\varepsilon)u_a^2 + \beta(\varepsilon)u_a^4$ | $\alpha(\varepsilon) = 347.8\varepsilon + 48.22$<br>$\beta(\varepsilon) = 3649$ | meV/Å$^2$<br>meV/Å$^4$ |
| Dipole-dipole | $\frac{2Z^*(\varepsilon)^2}{\epsilon_\infty(\varepsilon)} \sum_{a \neq b} Q_{ab,zz}(\varepsilon)u_{a,z}u_{b,z}$ | $\epsilon_\infty(\varepsilon) = 0.013\varepsilon^2 - 0.05\varepsilon + 6.68$<br>$Z^*(\varepsilon) = -0.005\varepsilon^2 + 0.113\varepsilon + 3.003$ | --<br>e |
| Short-range | $\frac{1}{2} \sum_{\lambda,<a,b>} J_\lambda(\varepsilon)_{ab}\vec{u}_a \cdot \vec{u}_b$ | $J_1(\varepsilon) = 1.368\varepsilon - 50.03$<br>$J_2 = -2.83$<br>$J_3 = 7.396$<br>$J_4 = 4.702$<br>$J_5 = 0.859$ | meV/Å$^2$ |

**Supplementary Table 1 |** The components of the effective lattice Hamiltonian for $BaFe_{12}O_{19}$.

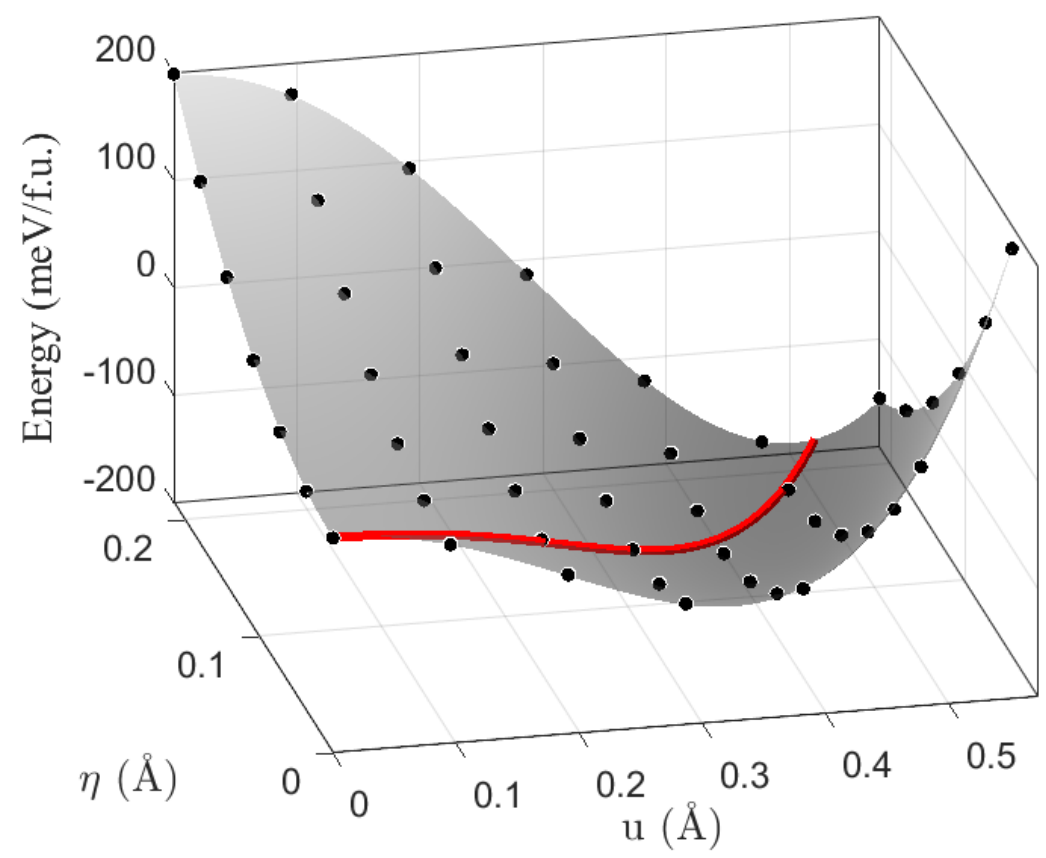

**Supplementary Fig. 11 |** An example of a fit to the DFT-calculated energy landscape (black dots) using a Landau expansion $E(u, \eta)$ (gray surface), at strain = –5%. The red curve shows the minimum energy at each $u$, where $\partial E/\partial\eta = 0$.

| Strain (%) | $\alpha$ (meV/Å$^2$) | $\beta$ (meV/Å$^4$) | RMSE (meV/f.u.) |
|---|---|---|---|
| -1 | -319.5 | 3867 | 0.07 |
| -1.5 | -488.3 | 3685 | 0.10 |
| -2 | -657.4 | 3660 | 0.15 |
| -2.5 | -846 | 3871 | 0.17 |
| -3 | -1012 | 3834 | 0.34 |
| -4 | -1360 | 3774 | 1.07 |
| -5 | -1690 | 3641 | 2.01 |

**Supplementary Table 2 |** Renormalized values of $\alpha$ and $\beta$ at different values of strains. The last column shows the root-mean-square error of fitting the energy landscape with the Landau expansion $E(u, \eta)$.

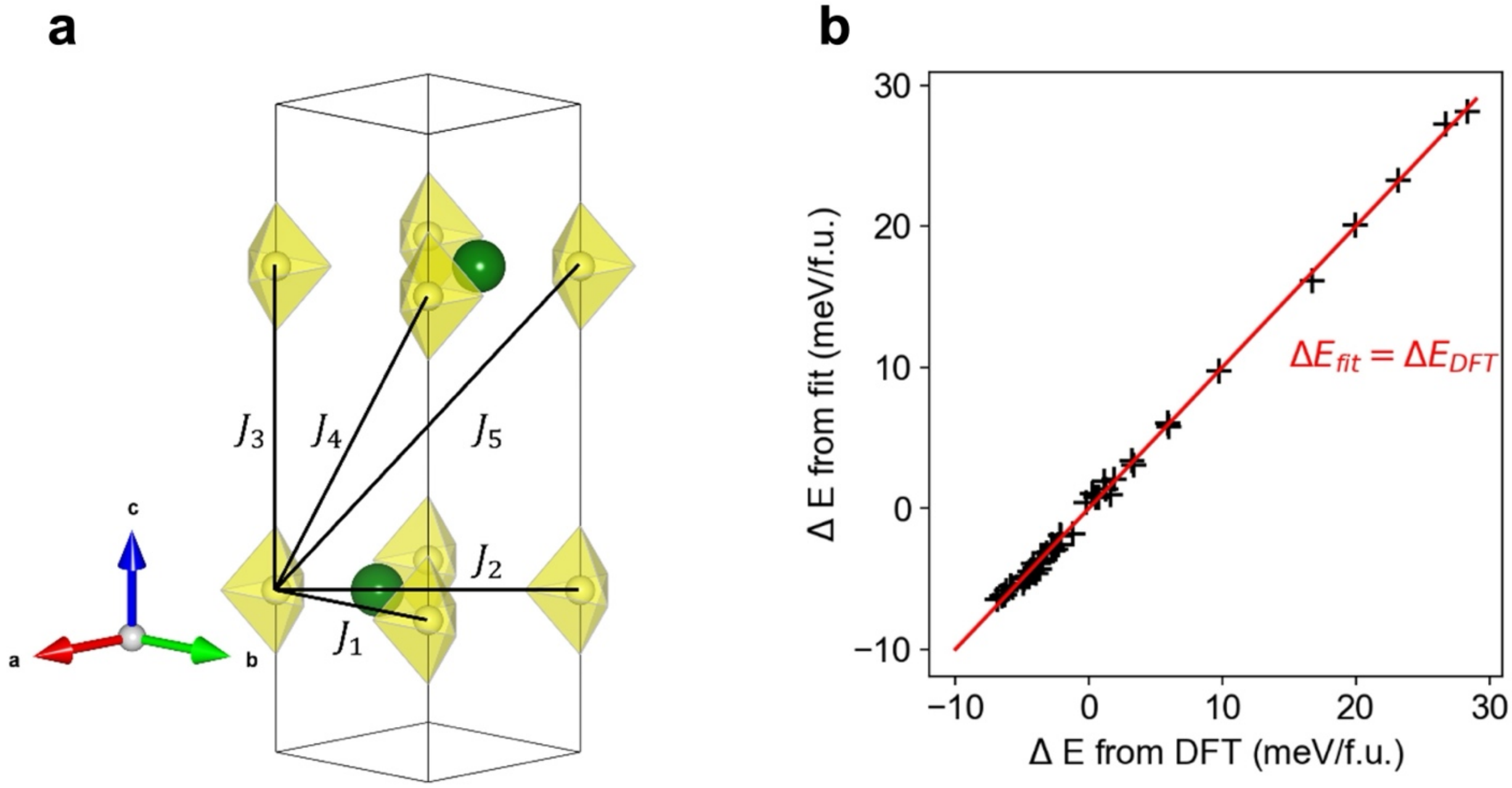

**Supplementary Fig. 12 | a,** Short-range interactions $J_1$ through $J_5$ in $BaFe_{12}O_{19}$. **b,** A comparison of shift in total energy from the ferroelectric (↑) configuration calculated from DFT and predicted by our fitting. All the data points are close to the red line, where fitted energy shifts match the DFT calculations exactly.

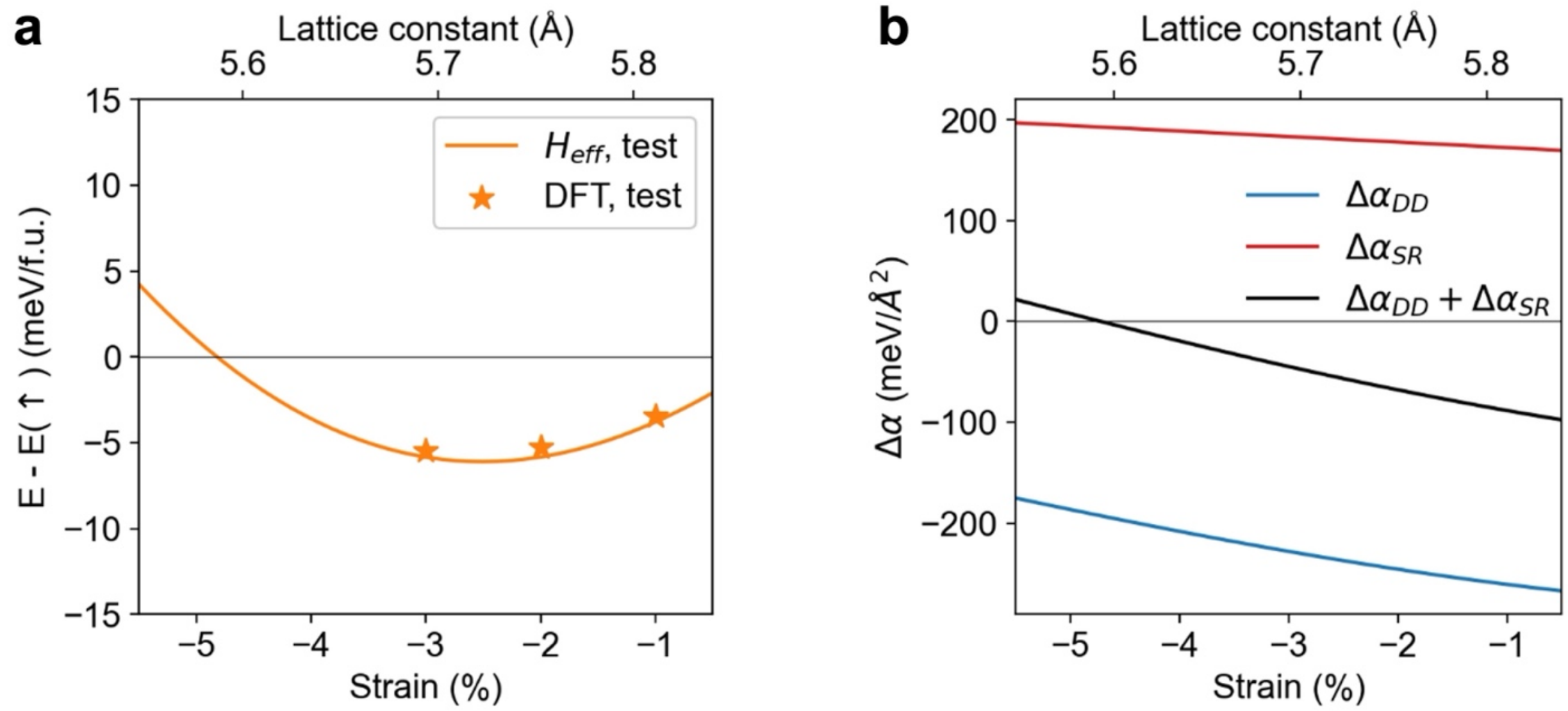

**Supplementary Fig. 13 | a,** Test points (star) using the 4 ↑ 2 ↓ electric dipole configuration in $BaFe_{12}O_{19}$ in the plane, and ferroelectric out of plane, which are not in the fitting dataset. Our effective Hamiltonian (solid curve) captures the energy of these test points well. **b,** Difference in the dipole-dipole coefficient $\Delta\alpha_{DD}(\varepsilon)$, short-range coefficient $\Delta\alpha_{SR}(\varepsilon)$, and their sum as a function of strain between the antiferroelectric ($\begin{smallmatrix}\uparrow & \downarrow \\ \uparrow & \downarrow\end{smallmatrix}$) and ferroelectric (↑) configurations.

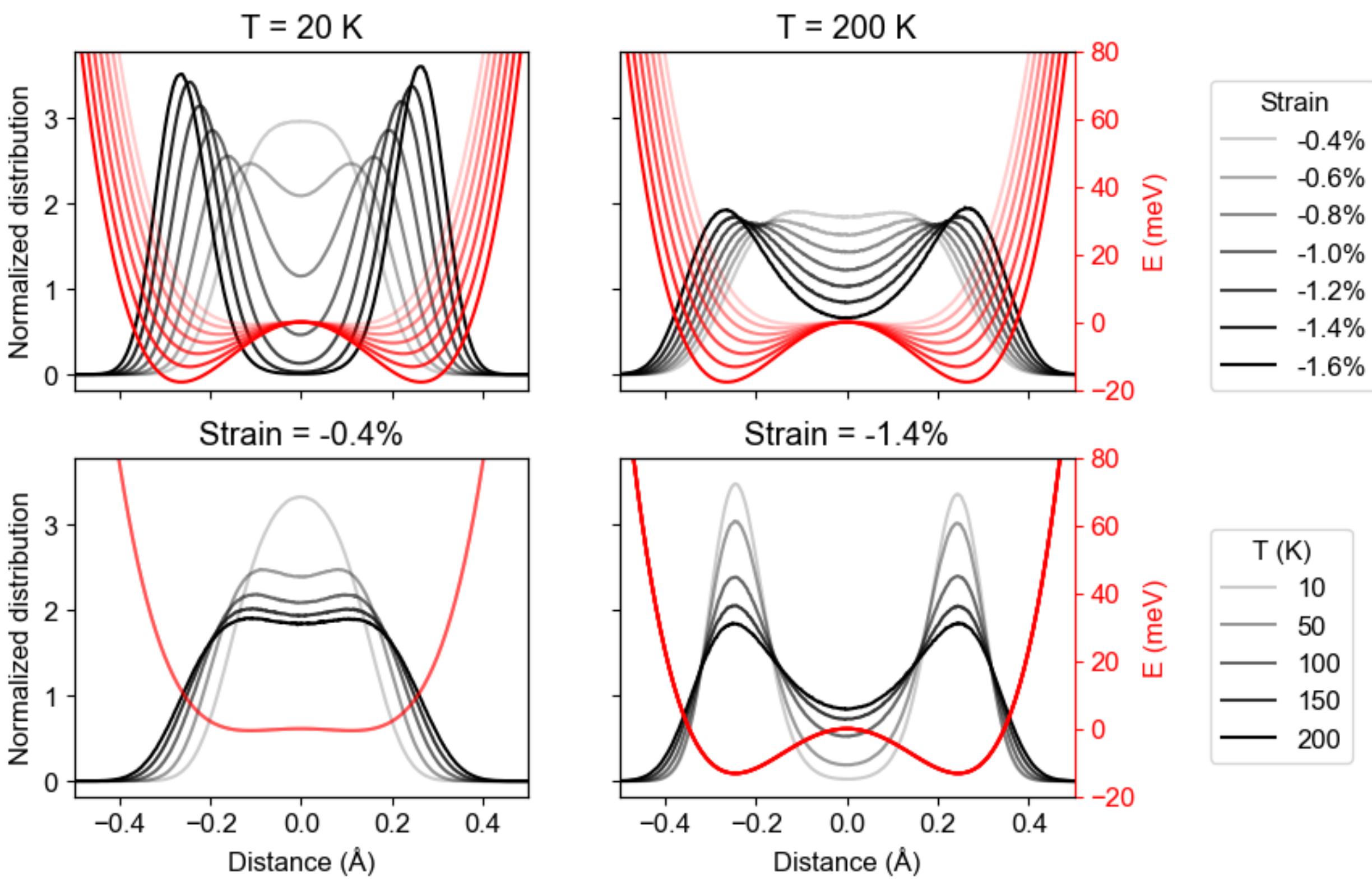

**Supplementary Fig. 14 |** Normalized distribution of iron in the trigonal bipyramid of $BaFe_{12}O_{19}$ obtained from PIMC, under various values of compressive strain and temperature. The red curves show the double-well potential in the ferroelectric configuration, i.e., $E(u, \varepsilon) = (347.8\varepsilon + 48.22)u^2 + 3649u^4$, which serves as an approximation to the local double-well potential without interactions.

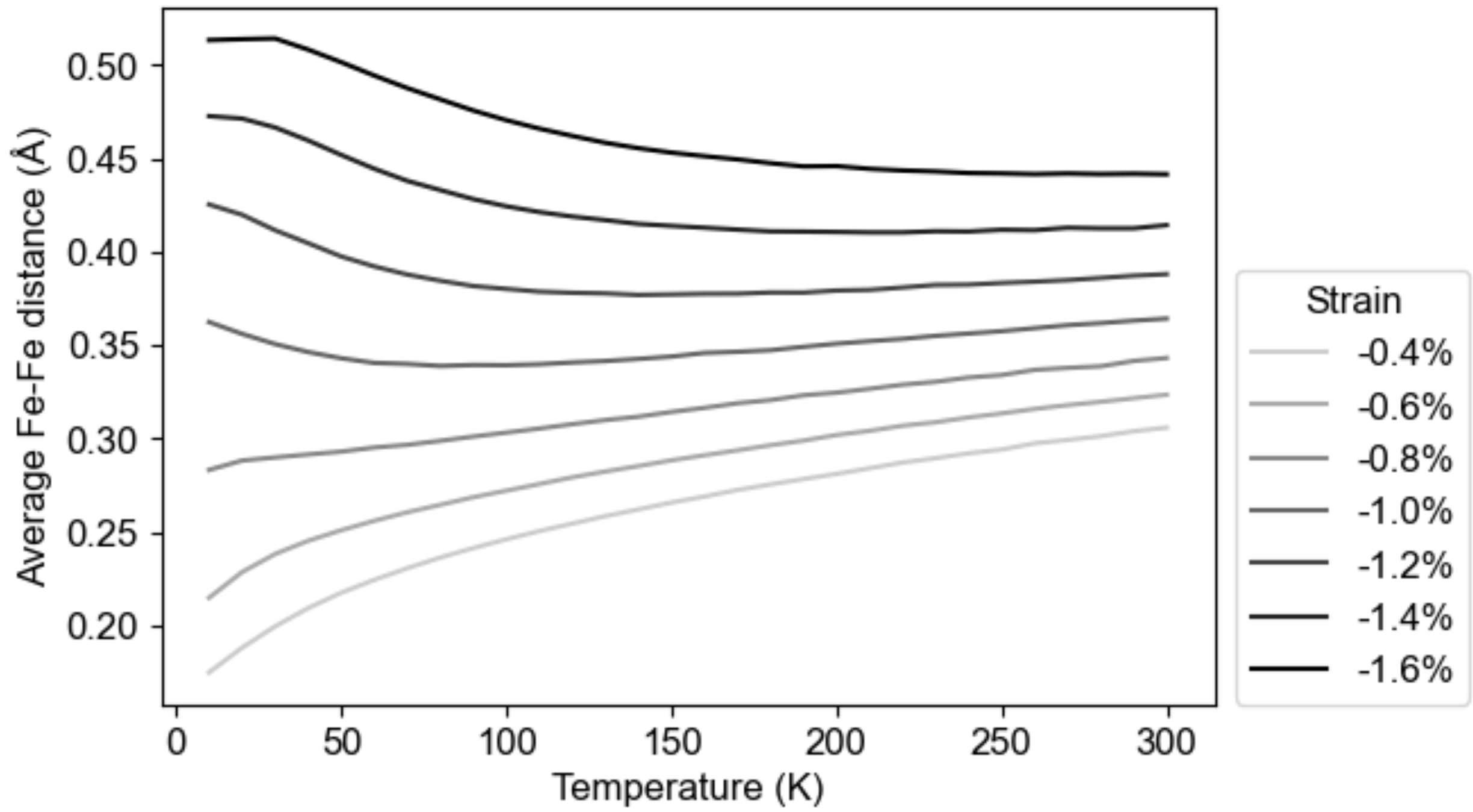

**Supplementary Fig. 15 |** Average Fe-Fe distance along the *c*-axis direction of the trigonal bipyramid site of $BaFe_{12}O_{19}$ as a function of temperature from 10 K to 300 K, and strain from –0.4% to –1.6%, calculated from the iron distribution $P(u)$ as $2(\sum P(u)|u|)/\sum P(u)$.

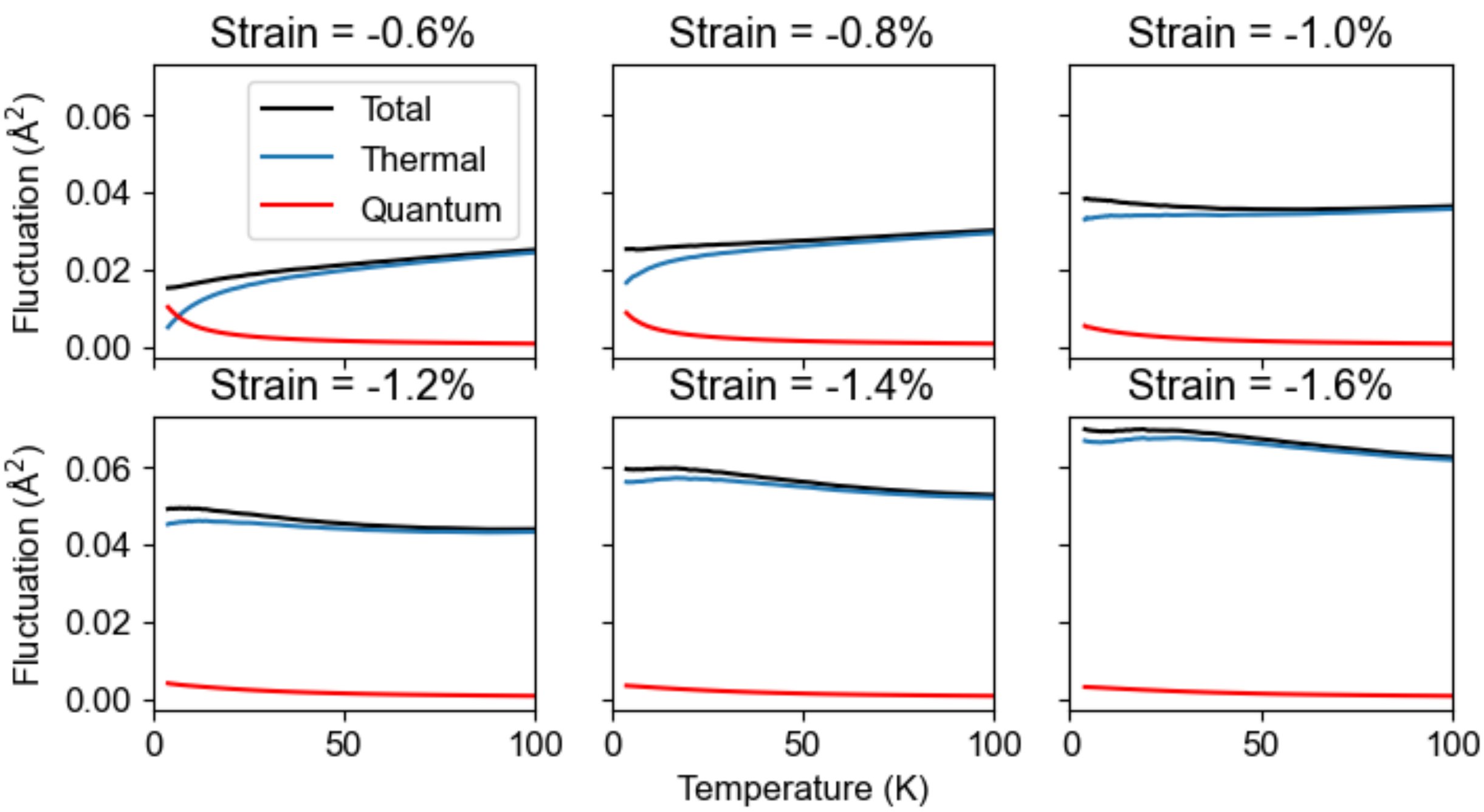

**Supplementary Fig. 16 |** Total fluctuation of strained $BaFe_{12}O_{19}$, which is split into thermal and quantum parts, vs. temperature from 4 K to 100 K, and strain between –0.6% and 1.6%.

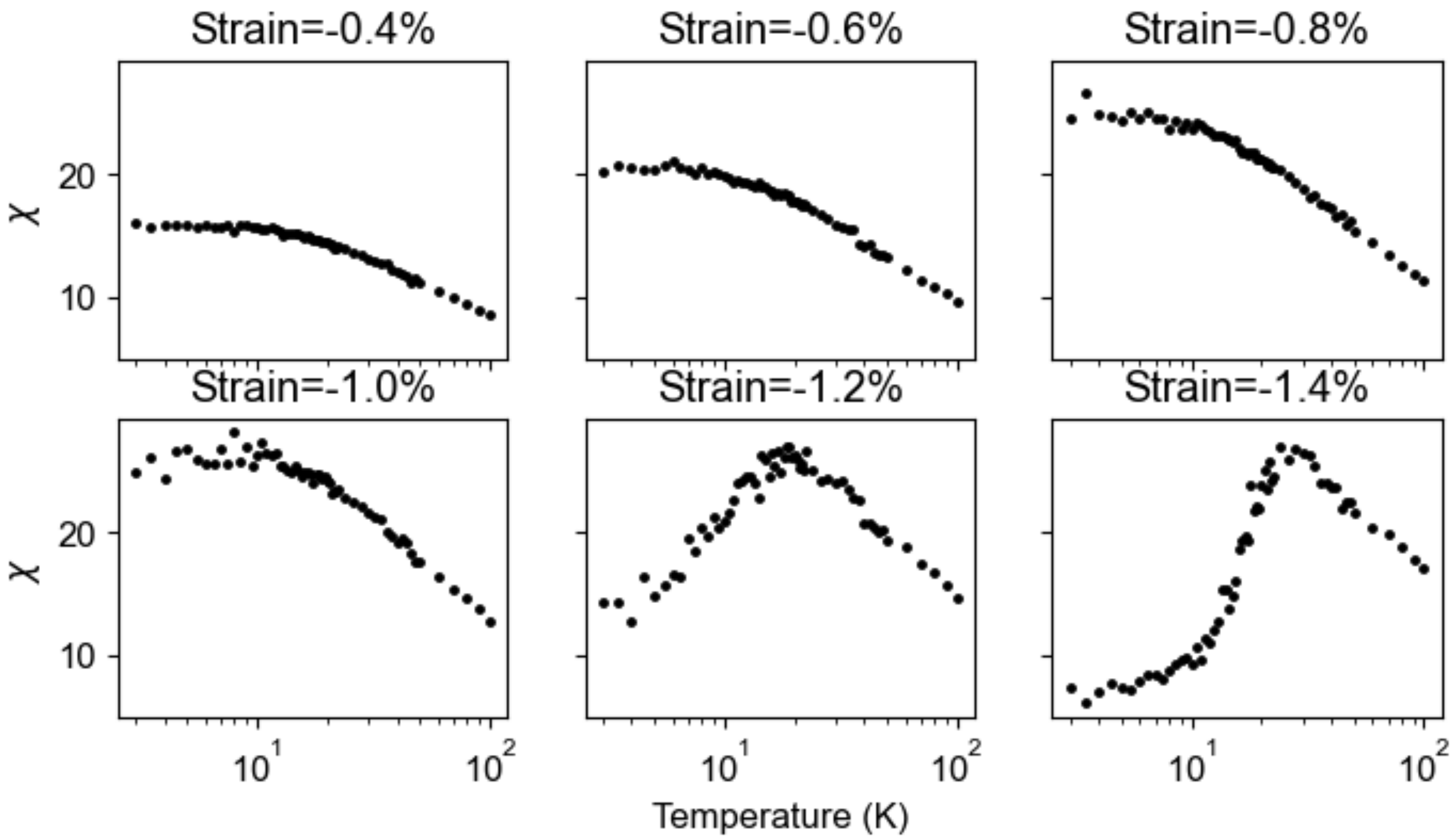

**Supplementary Fig. 17 |** PIMC simulated dielectric susceptibility of strained $BaFe_{12}O_{19}$ between 4 K and 100 K. Note the *x*-axis is plotted on a log scale.

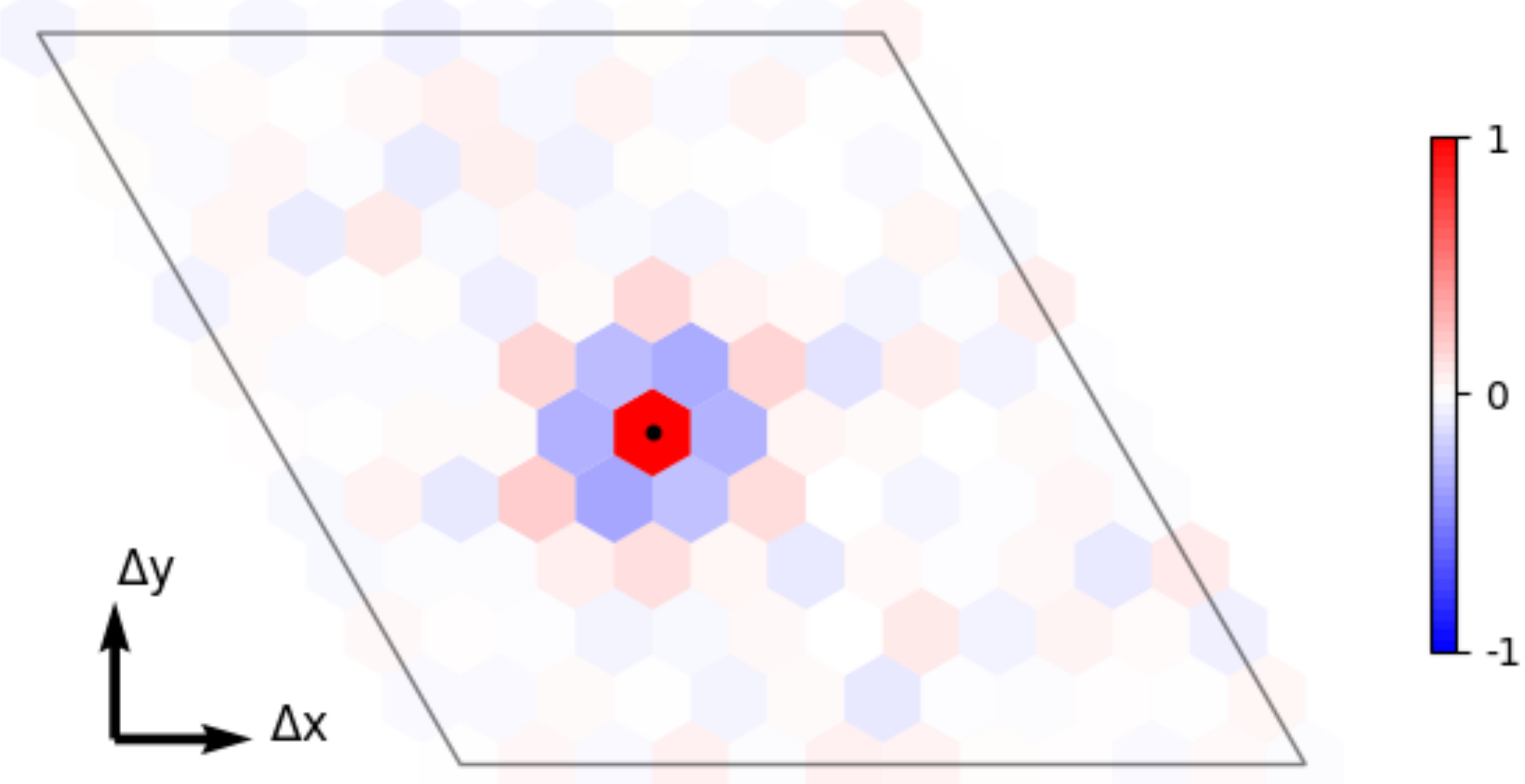

**Supplementary Fig. 18 |** Real-space dipole correlation map in plane. $S(\Delta x, \Delta y)$ indicates the degree of (anti)-alignment between a dipole at a distance $(\Delta x, \Delta y)$ and (0,0), the location of the black dot.